\documentclass[11pt,a4paper]{article}
\usepackage[sc,noBBpl]{mathpazo}
\usepackage[mathscr]{eucal}
\usepackage{amsmath,amsfonts,amssymb,amsthm}
\usepackage{graphicx}

\newcommand\bI{{\mathbf I}}

\newcommand\cS{{\mathcal S}}

\newcommand\scA{{\mathscr A}} 
 
\newcommand\scC{{\mathscr C}} 
\newcommand\scD{{\mathscr D}}

\newcommand\scI{{\mathscr I}}

\newcommand\scL{{\mathscr L}} 
\newcommand\scM{{\mathscr M}} 

\newcommand\scO{{\mathscr O}}

\newcommand\scV{{\mathscr V}}

\newcommand\mvector{\boldsymbol}

\newcommand\va{\mvector{a}}
\newcommand\vb{\mvector{b}}
\newcommand\vc{\mvector{c}}
\newcommand\vd{\mvector{d}}

\newcommand\vf{\mvector{f}}

\newcommand\vh{\mvector{h}}

\newcommand\vp{\mvector{p}}
\newcommand\vq{\mvector{q}}

\newcommand\vs{\mvector{s}}

\newcommand\vx{\mvector{x}}
\newcommand\vy{\mvector{y}}
\newcommand\vz{\mvector{z}}
\newcommand\vA{\mvector{A}}
\newcommand\vB{\mvector{B}}
\newcommand\vC{\mvector{C}}

\newcommand\vE{\mvector{E}}
\newcommand\vF{\mvector{F}}

\newcommand\vH{\mvector{H}}

\newcommand\vK{\mvector{K}}
\newcommand\vL{\mvector{L}}

\newcommand\vN{\mvector{N}}

\newcommand\vP{\mvector{P}}
\newcommand\vQ{\mvector{Q}}

\newcommand\vS{\mvector{S}}

\newcommand\vU{\mvector{U}}

\newcommand\vW{\mvector{W}}

\newcommand\vLambda{\mvector{\Lambda}}

\newcommand\vvarphi{\mvector{\varphi}}

\newcommand\vzero{\mvector{0}}

\newcommand\field{\mathbb}

\newcommand\R{\field{R}}

\newcommand\C{\field{C}}
\newcommand\Z{\field{Z}}
\newcommand\M{\field{M}}
\newcommand\N{\field{N}}
\newcommand\Q{\field{Q}}

\newcommand\bbP{\mathbb{P}}

\newcommand\rank{\operatorname{rank}}
\newcommand\diag{\operatorname{diag}}

\newcommand\tr{\operatorname{Tr}}
\newcommand\res{\operatorname{res}}
\newcommand\result{\operatorname{Res}}
 
 \newcommand\spectr{\operatorname{spectr}}

\newcommand\grad{\operatorname{grad}}

\newcommand\card{\operatorname{card}}

\newcommand\rmd{\mathrm{d}\mspace{1mu}}

\newcommand\CP{\ensuremath{\C\bbP}}
\newcommand\rmi{\mathrm{i}\mspace{1mu}}

\newcommand\Dt{\frac{\mathrm{d}\phantom{t} }{\mathrm{d}\mspace{1mu} t}}

\newcommand\pder[2]{\dfrac{\partial #1 }{\partial #2}} 
\newcommand\bfi[1]{{\bfseries\itshape{#1}}}

\newcommand\mtext[1]{\quad\text{#1}\quad}

\newcommand\defset[2]{\left\{{#1}\;\vert \;\; {#2} \,\right\}}
\newcommand\deftuple[2]{\left({#1}\;\vert \;\; {#2} \,\right)}

\usepackage[square]{natbib}
\theoremstyle{plain}

\newtheorem{theorem}{Theorem}
\newtheorem{lemma}{Lemma}
\newtheorem{proposition}{Proposition}

\theoremstyle{definition}

\newtheoremstyle{note}{\topsep}{\topsep}{\slshape}{}{\scshape}{}{ }{}
\theoremstyle{note}

\newtheorem{conjecture}{Conjecture}
\newtheorem{remark}{Remark}

\numberwithin{equation}{section}
\numberwithin{theorem}{section}
\numberwithin{lemma}{section}
\numberwithin{proposition}{section}
\numberwithin{remark}{section}
\numberwithin{example}{section}
\numberwithin{assumption}{section}
\numberwithin{conjecture}{section}

\newcommand{\adots}{\mbox{\setlength{\unitlength}{1pt}
                          \begin{picture}(8,7)\put(-4,1){.}\put(0,4){.}
                          \put(4,7){.}\end{picture}}}

\numberwithin{figure}{section}
\numberwithin{table}{section}

\begin{document}
\thispagestyle{empty}
\vspace*{1em}
\begin{center}
\LARGE\textbf{Darboux points and integrability of homogeneous Hamiltonian
systems with
three and more degrees of freedom. Nongeneric cases}
\end{center}
\vspace*{0.5em}
\begin{center}
\large  Maria Przybylska
\end{center}
\vspace{2em}
\hspace*{2em}\begin{minipage}{0.8\textwidth}
\small
Toru\'n Centre for Astronomy,
  N.~Copernicus University, \\
  Gagarina 11, PL-87--100 Toru\'n, Poland, \\
  (e-mail: Maria.Przybylska@astri.uni.torun.pl)
\end{minipage}\\[2.5em]
{\small \textbf{Abstract.}
In this paper  the problem of  classification of  integrable natural  Hamiltonian systems with $n$ degrees of freedom given by
 a Hamilton function which is the sum of the standard kinetic energy and a homogeneous polynomial potential
 $V$  of degree $k>2$ is investigated. It is assumed that the potential is  not generic. Except for some particular cases  a potential $V$ is not generic, if it admits  a nonzero solution of equation $V'(\vd)=0$. The existence of such solution gives very strong integrability obstructions obtained in the frame  of the Morales-Ramis theory. This theory gives also additional integrability obstructions  which have the form of restrictions imposed on  the eigenvalues $(\lambda_1,\ldots,\lambda_n)$ of the Hessian matrix $V''(\vd)$  calculated at a non-zero $\vd\in\C^n$ satisfying
$V'(\vd)=\vd$.
In our previous work we showed that  for generic  potentials  some universal relations between $(\lambda_1,\ldots,\lambda_{n})$ calculated at various solutions of $V'(\vd)=\vd$ exist. These relations allow to prove
that the number  of potentials satisfying the necessary conditions for the integrability is finite.
The main aim of this paper was to show that  relations of such forms also exist  for nongeneric potentials. We show their existence and derive them for case $n=k=3$ applying
 the multivariable residue calculus.
We demonstrate the
strength of the obtained results analysing in details the nongeneric cases for $n=k=3$. Our analysis cover all  the possibilities and we distinguish those cases where known methods are too weak to decide if the potential is integrable or not.
Moreover, for $n=k=3$ thanks to this analysis a three-parameter family of potentials integrable or super-integrable with additional polynomial first integrals which seemingly can be of  an
arbitrarily high degree with respect to the momenta  was distinguished.

For an arbitrary $n>2$ and $k>2$ we show how to distinguish different types of nongeneric potentials and we analyse their integrability properties.
\\
{\bf MSC2000 numbers:} 37J30, 70H07, 37J35, 34M35.\\
{\bf Key words:} Integrability, Hamiltonian systems, Homogeneous potentials,  Differential Galois group.
}

\section{Introduction}
In this paper we consider natural
Hamiltonian systems given by the following Hamiltonian
\begin{equation}
H=\dfrac{1}{2}\vp^{T}\vp+V(\vq), \qquad \vq,\vp\in\C^n,
\label{eq:ham}
\end{equation}
where $V(\vq)$ is a homogeneous polynomial of degree $k>2$. The canonical equations
corresponding to the above Hamiltonian  have the form
\begin{equation}
\label{eq:eqham}
\Dt \vq= \vp, \qquad \Dt \vp =-V'(\vq),
\end{equation}
where $V'(\vq):=\grad V(\vq)$ denotes the gradient of $V(\vq)$. We say that a potential $V$ is
integrable if the above canonical equations are integrable in the Liouville
sense.

This paper is a continuation of our previous work \cite{mp:08::d} where the following classification problem was formulated.
\begin{quotation}
\noindent
{\it Give a complete list of integrable homogeneous polynomial potentials for given $k>2$ and $n\geq 2$. In other words:
formulate  necessary and sufficient conditions for the integrability of homogeneous polynomial potentials.}
\end{quotation}
For two  degrees of freedom and  for small $k$ this problem was analysed and solved  in \cite{mp:04::d,mp:05::c}.  However, methods from those papers do not have direct  extensions  for $n>2$ degrees of freedom.   As it was shown in~\cite{mp:08::d}, the problem in higher dimensions is  difficult but nevertheless it is tractable with  more advanced methods of algebraic geometry and the multivariable residue calculus. All these techniques are described in \cite{mp:08::d}. In that  paper all  integrable potentials with $n=k=3$  and satisfying certain `genericity' assumption were found.   Later we  give a precise definition of  a generic homogeneous polynomial potential, but, independently what the genericity means, it must be underlined that
the integrability is an extremely exceptional phenomenon, and this is why  we cannot exclude from our considerations certain classes of potentials.

The aim of this paper is to investigate the integrability properties of nongeneric homogeneous polynomial potentials.  It appears that among them we can find also integrable ones.

The plan of this paper is the following. In the next section~\ref{sec:darbob} we define more precisely the notion of the Darboux point that is crucial in the whole integrability analysis  and recapitulate all integrability obstructions caused by the presence of  such points. In Section~\ref{sec:generaln} some general results concerning any number $n$ of degrees of freedom  are presented. The rest of this paper constitutes Section~\ref{sec:threeclass} containing the integrability analysis for $n=k=3$ in nongeneric cases ordered by the number of proper Darboux points. For convenience of the reader the main results obtained in this section are summarised just at its beginning.

\begin{remark}
\label{rem:1}
 As in~\cite{mp:08::d} we divide all potentials into equivalent classes.  %
We say that $V$ and $\widetilde V$ are equivalent if there exists a matrix $\vA\in
\mathrm{PO}(n,\C)$
such that $\widetilde V(\vq)=V_{\vA}(\vq):=V(\vA\vq)$.   Here  $\mathrm{PO}(n,\C)$ denotes
 the complex projective
orthogonal group, defined by
\begin{equation}
\mathrm{PO}(n,\C)=\{\vA\in\mathrm{GL}(n,\C),\ |\ \vA\vA^T=\alpha \vE_n,\;
\alpha\in\C^\star\},
\end{equation}
and $\vE_n$ is $n$-dimensional identity matrix.
Later a
potential means a class of equivalent potentials in the above sense.
\end{remark}
\begin{remark}
Let us consider a following Hamiltonian function
\[
 H=\dfrac{1}{2}\vp^T\vK\vp+V(\vq),
\]
where  $\vK^T=\vK$ is semi-simple, and $\det \vK\neq 0$.
It is easy to show that we can find a linear symplectic transformation
\[
 \vq = \vA\vQ, \quad \vp=\vA^{-1}\vP,
\]
such that in new variables the Hamiltonian reads
\[
 H=\dfrac{1}{2}\vP^T\vP+W(\vQ),
\]
where $W(\vQ):=V(\vA\vQ)$. This shows  that the restriction of the kinetic energy in \eqref{eq:ham} to $T=\frac{1}{2}\vp^T\vp$ is in fact equivalent to  the assumption that $T$ is a quadratic form in the momenta with constant coefficients.
\end{remark}
\begin{remark}
 Many integrable potentials found in this paper as well as in \cite{mp:08::d} have complex coefficients.  The question is if such potentials can be equivalent to real ones. In other words, whether exists  $A\in\mathrm{PO}(n,\C)$ such that for a given $V\in\C[\vq]$ we have $V_{\vA}\in\R[\vq]$. It seems that this question is difficult even for $n=3$. Nevertheless, for all integrable potentials which are given in this paper as well as in \cite{mp:08::d} we can find a linear canonical change of variables which transforms Hamiltonian~\eqref{eq:ham} with $V\in\C[\vq]$ into
\[
 H=\dfrac{1}{2}\vp^T\vK\vp+\widetilde V(\vq),
\]
where $\widetilde V\in\R[\vq]$, and
\begin{equation}
 \vK=\diag(\varepsilon_1, \ldots, \varepsilon_n),  \qquad \varepsilon_i\in\{-1,1\}.
\end{equation}
\end{remark}

\section{Darboux points and obstructions to the integrability}
\label{sec:darbob}

In this section we collected all known integrability obstructions for homogeneous polynomial potentials. There are two types of them.  The first  ones are obtained by an application of the  Morales-Ramis theory, see \cite{Morales:01::b1,Morales:99::c,Audin:01::,Audin:02::,Morales:07::} that formulates the necessary integrability conditions in terms of properties of the differential Galois group of variational equations along a certain non-stationary particular solution.
In the case  of canonical equations~\eqref{eq:eqham} we  look for a particular solution of the form
\begin{equation}
 \vq(t) = \varphi(t)\vd, \qquad \vp(t) =\dot \varphi(t)\vd,
\label{eq:ppart}
\end{equation}
where $\vd\in\C^n$ is a non-zero vector, and $ \varphi(t)$  is a scalar function satisfying $\ddot \varphi =-  \varphi^{k-1}$.  Then, it appears, that all the properties of differential Galois group of variational equations along this particular solution can be expressed in terms
 of eigenvalues of Hessian matrix $V''(\vd)$.

The other type of integrability obstructions for homogeneous polynomial potentials can be obtain by a certain kind of global analysis. It is easy to see that if ~\eqref{eq:ppart} is a solution~\eqref{eq:eqham}, then
$\vd$ is a solution of the following equation
\begin{equation}
\label{eq:vdp}
 V'(\vd)=\gamma \vd, \mtext{where} \gamma\in\C^\star.
\end{equation}
Each solution $\vd\neq\vzero$ of the above algebraic equations gives obstructions for the integrability expressed in terms of eigenvalues of $V''(\vd)$.    It can be shown that between all the eigenvalues taken at all possible solutions of~\eqref{eq:vdp}  certain relations  exist which  give very strong restrictions to the integrability if we combine them with results obtained from the Morales-Ramis theory.
We see that solutions of \eqref{eq:vdp} play a crucial  role. We start with
some  definitions which give a proper geometric and algebraic framework to study the set of these solutions.

Let us recall that a point in the $m$ dimensional complex projective space $\CP^m$ is specified by
its homogeneous coordinates $[\vz]=[z_0: \cdots:z_m]$, where
$\vz=(z_0,\ldots,z_m)\in\C^{m+1}\setminus\{\vzero\}$; moreover, $\vz\neq\vzero$, and $\lambda\vz$ with $\lambda\in\C^\star$ define the same point $[\vz]=[\lambda\vz]$ in $\CP^m$. Let
\begin{equation}
 U_i:=\defset{[z_0: \cdots:z_m]\in\CP^m}{z_i\neq 0}\mtext{for}i=0,\ldots,m,
\end{equation}
 then
\begin{equation}
 \CP^m=\bigcup_{i=0}^m U_i,
\end{equation}
and we have natural coordinate maps
\begin{equation*}
 \theta_i:\CP^m \supset U_i\rightarrow \C^m ,\qquad
  \theta_i([\vz])=(x_1,\ldots,x_m),
\end{equation*}
where
\begin{equation}
 (x_1,\ldots,x_m)=\left(\frac{z_1}{z_i}, \ldots,
 \frac{z_{i-1}}{z_i}, \frac{z_{i+1}}{z_i}, \ldots, \frac{z_{m}}{z_i}\right).
\end{equation}
Each $U_i$ is homeomorphic to $\C^m$. It is easy to check that charts
$(U_i,\theta_i)$, $i=0,\ldots, m$ form an atlas which makes $\CP^m$ an
holomorphic $m$-dimensional manifold.

Let $V$ be a homogeneous polynomial potential of degree $k>2$, i.e.,
$V\in\C_k[\vq]$, and $\vd\in\C^n\setminus\{\vzero\}$. We say that $[\vd]\in\CP^{n-1}$  is   \bfi{a Darboux point} of $V$ iff
\begin{equation}
 \vd\wedge V'(\vd)=\vzero , \qquad \vd\neq \vzero.
\end{equation}
The set ${\scD}(V) \subset \CP^{n-1}$ of all Darboux points of a potential $V$
is a projective algebraic set. In fact, ${\scD}(V)$ is the zero locus in $\CP^{n-1}$ of
homogeneous polynomials $R_{i,j}\in\C_k[\vq]$ which are components of $\vq\wedge
V'(\vq)$, i.e.
\begin{equation}
\label{eq:Rij}
 R_{i,j}:= q_i\pder{V}{q_j}-q_j \pder{V}{q_i}, \mtext{where} 1\leq i<j\leq n.
\end{equation}

We say that a Darboux point $[\vd]\in {\scD}(V)$ is \bfi{a proper Darboux point} of $V$, iff
$V'(\vd)\neq \vzero$. The set of all proper Darboux points of $V$ is denoted by
${\scD}^\star(V)$. If  $[\vd]\in {\scD}(V)\setminus {\scD}^\star(V)$, then  $[\vd]$ is called \bfi{an improper
Darboux point} of potential $V$.

If  $[\vd]$ is a Darboux point of $V$, then it is called an \bfi{isotropic Darboux point}, iff
 \begin{equation}
 \label{eq:iso}
 d_1^2+\cdots+d_n^2=0.
 \end{equation}

We say that homogeneous polynomial potential $V$ of degree $k>0$ is \bfi{generic} if  all its Darboux points are proper and simple. In this case it has exactly
\[
 D(n,k)=\dfrac{(k-1)^n-1}{k-2},
\]
Darboux points.
\begin{remark}
 The set of all homogeneous polynomials of degree $k$ in $n$ variables is a $\C$-vector space of dimension
$\binom{n+k-1}{k}$.
\end{remark}

If we choose the affine chart $(U_1, \theta_1)$ on $\CP^{n-1}$, then Darboux points located on $U_1$ are characterised in the following way.
\begin{lemma}
\label{lem:aff}
On the affine chart $(U_1, \theta_1)$ we have
\begin{equation}
\label{eq:dvu1}
\theta_1(\scD(V)\cap U_1)=\scV(g_1,\dots,g_{n-1}),
\end{equation}
where polynomials $g_1,\ldots,g_{n-1}\in\C[\widetilde \vx]$ are given by
\begin{equation}
\label{eq:g0}
v(\widetilde\vx):=V(1,x_1,\ldots, x_{n-1}), \quad g_0:= k v -\sum_{i=1}^{n-1}
 x_i\pder{v}{x_i},
\end{equation}
and
\begin{equation}
\label{eq:hg}
g_i := \pder{v}{x_i}-x_i g_0, \mtext{for}i=1, \ldots, n-1.
\end{equation}
Moreover, $[\vd]\in \scD(V)\cap U_1$ is an improper Darboux point iff its affine
coordinates $\widetilde \va := \theta_1([\vd])$ satisfy $g_0( \widetilde
\va)=0$.
\end{lemma}

 The presence of a proper Darboux point yields quite strong integrability obstructions. Let us assume that $V$ possesses a proper Darboux point $[\vd]$. We assume that $V''(\vd)$ is diagonalisable with eigenvalues  $(\lambda_1,\ldots, \lambda_n)$. Vector $\vd$ is an eigenvector of $V''(\vd)$ with eigenvalue $\lambda_n= k-1$,  and we  called this eigenvalue trivial.
Morales and Ramis have proved in \cite{Morales:01::a}, see also \cite{Morales:99::c}, the following result.
\begin{theorem}
\label{thm:MoRa}
If Hamiltonian system~\eqref{eq:eqham} with polynomial homogeneous
potential $V(\vq)$ of degree $k>2$ is meromorphically integrable in
the Liouville sense, then for a proper Darboux point the values of
$(k,\lambda_i)$ for $i=1,\ldots,n$ belong to the following list
\begin{equation}
\label{eq:tabMoRa}
\begin{tabular}{clcl}
1.& $\left( k, p + \dfrac{k}{2}p(p-1)\right)$,&2.& $\left(k,\dfrac 1
{2}\left[\dfrac {k-1} {k}+p(p+1)k\right]\right)$, \\[1em]
3.& $\left(3,-\dfrac 1 {24}+\dfrac 1 {6}\left( 1 +3p\right)^2\right)$, &
4.& $\left(3,-\dfrac 1 {24}+\dfrac 3 {32}\left(  1  +4p\right)^2\right)$,
\\[1em]
5.& $\left(3,-\dfrac 1 {24}+\dfrac 3 {50}\left(  1  +5p\right)^2\right)$,&
6.& $\left(3,-\dfrac 1 {24}+\dfrac{3}{50}\left(2 +5p\right)^2\right)$,\\[1em]
7.& $ \left(4,-\dfrac 1 8 +\dfrac{2}{9} \left( 1+ 3p\right)^2\right)$,&
8.& $\left(5,-\dfrac 9 {40}+\dfrac 5 {18}\left(1+ 3p\right)^2\right)$,\\[1em]
9.& $\left(5,-\dfrac 9 {40}+\dfrac 1 {10}\left(2+5p\right)^2\right)$,&
 &
\end{tabular}
\end{equation}
where $p$ is an integer.
\end{theorem}
We denote by $\mathscr{M}_k$ a subset of rational numbers $\lambda$ specified by items of
the table in the above theorem for a given $k$.
\begin{remark}
 Let us note that the eigenvalues of $V''(\vd)$ depend on the representative of the Darboux point $[\vd]$. Thus instead of $\lambda_i$ one should use $\lambda_i/\lambda_n$ as these quantities that are well defined functions of the Darboux point. However, because of long tradition we accept the following convention. If $[\vd]\in\CP^{n-1}$ is a proper Darboux point of $V$, then we always choose its representative $\vd\in\C^n$ in such a way that it satisfies
 $V'(\vd)=\vd$.
\end{remark}
 \begin{remark}
 \label{rem:nd}
It was explained in \cite{Duval:08::}  that the assumption that $V''(\vd)$ is
diagonalisable is irrelevant. That is, the necessary conditions for the
integrability are the same: if the potential is integrable, then each
$\lambda\in\spectr V''(\vd)$ must belong to appropriate items of the above list. Additionally, if $V''(\vd)$ is
not diagonalisable, then new obstacles for the integrability appear. Namely, if the
Jordan form of $V''(\vd)$ has a block
\begin{equation*}
J_3(\lambda):=
\begin{bmatrix}
\lambda &1&0\\
0&\lambda &1\\
0&0&\lambda
\end{bmatrix},
\end{equation*}
 then the system is not integrable. Moreover, if the Jordan form of $V''(\vd)$
 has a two dimensional block $J_2(\lambda)$, and $\lambda$ belongs to the first
 item of table~\eqref{eq:tabMoRa}, then the system is not integrable. This fact was proved in~\cite{Duval:08::}.

In other words, the presence of a proper Darboux point $\vd$ for that the Jordan form of $V''(\vd)$ has a block of dimension greather than two or
degree two with $\lambda_i$ that belongs to the first
 item of table~\eqref{eq:tabMoRa} implies immediately the nonintegrability of the potential.
\end{remark}
It appeared quite recently, see \cite{mp:08::d}, that improper Darboux points also give very strong integrability obstructions.
\begin{theorem}
\label{thm:im}
Assume that a homogeneous potential $V\in\C_k[\vq]$ of degree $k>2$ admits an
improper Darboux point $[\vd]\in\CP^{n-1}$. If $V$ is integrable with rational
first integrals, then matrix $V''(\vd)$ is nilpotent, i.e., all its eigenvalues
vanish.
\end{theorem}

Eigenvalues of $V''(\vd)$ taken over all proper  Darboux points are not arbitrary.   There are some  relations between them. Moreover, these relations  have the same form for  \bfi{an arbitrary} potential of a fixed degree $k$ and  satisfying certain genericity assumptions.  We say that these relations are universal as they do not depend on the values of the potential coefficients nor on its integrability properties.
For their description we  define \bfi{the spectrum of proper Darboux point} $[\vd]$  as  the $(n-1)$-tuple  $\vLambda(\vd)=(\Lambda_1(\vd),\ldots,
\Lambda_{n-1}(\vd))$, where $\Lambda_i(\vd):= \lambda_i(\vd)-1$ for $i=1,\ldots,n-1$.
For $0\leq r \leq n-1$ we denote by  $\tau_r$
the elementary symmetric polynomials in $(n-1)$ variables  of degree $r$, i.e.,
\[
\tau_r(\vx):=\tau_r(x_1,\ldots,x_{n-1})=\sum_{1\leq i_1<\cdots<i_r\leq
n-1}\prod_{s=1}^r x_{i_s}, \qquad 1\leq r\leq n-1,
\]
and $\tau_0(\vx):=1$. Then one can prove the following theorem \cite{mp:07::a,mp:08::d}.
\begin{theorem}
\label{thm:1}
  Let $V\in\C_k[\vq]$ be a homogeneous potential of integer degree $k>2$,  and let all its
Darboux points be proper and simple.
Then relations
\begin{equation}
 \sum_{[\vd]\in \mathscr{D}^\star(V)}
\frac{\tau_1(\vLambda(\vd))^r}{\tau_{n-1}(\vLambda(\vd)
)}=(-1)^{n-1}(-n-(k-2))^r,
\label{eq:rkoj}
\end{equation}
and
\begin{equation}
 \sum_{[\vd]\in \mathscr{D}^\star(V)}
\frac{\tau_r(\vLambda(\vd))}{\tau_{n-1}(\vLambda(\vd)
)}= (-1)^{r+n-1}\sum_{i=0}^{r}\binom{n-i-1}{r-i}(k-1)^{i},
\label{eq:rtau}
\end{equation}
for $r=0,\ldots,n-1$
are satisfied.
\end{theorem}
There exist generalisations of Theorem~\ref{thm:1} with weaker assumptions in the following cases
\begin{itemize}
 \item  $k=3$ and all  proper Darboux points $V$ are simple,
\item  $k>2$, all proper Darboux points of $V$ are simple and all improper Darboux points are minimally degenerated, for the definition of this notion see \cite{mp:08::d}.
\end{itemize}
In both these cases  relations~\eqref{eq:rkoj} and \eqref{eq:rtau} with $r=0,\ldots,n-2$, are satisfied.

The existence of these relations enables to prove the finiteness theorem, see Theorem~3.2 in~\cite{mp:08::d}.
To describe this result  and for later use let us recall some notions from~\cite{mp:08::d}.

 Let $\scC_m$ denote the set of all  unordered tuples  $\vLambda=(\Lambda_1, \ldots,
\Lambda_{m})$, where $\Lambda_i\in\C$ for $i=1,\ldots, m$. For $M>0$, the symbol
$\scC_m^M$ denotes the set of all unordered tuples
$(\vLambda_1,\ldots,\vLambda_M)$, where $\vLambda_i\in\scC_m$, for
$i=1,\ldots,M$.

 We fix $k>2$ and $n\geq 2$, and say that
 a
tuple  $\vLambda\in\scC_{n-1}$ is \bfi{admissible} iff
$\lambda_i=\Lambda_i+1\in\scM_k$ for $i=1,\ldots, n-1$. In other words, $\vLambda$ is admissible
iff $\Lambda_i+1$  belongs to
items, appropriate for a given $k$,  in the table of the Morales-Ramis
Theorem~\ref{thm:MoRa}, for $i=1,\ldots, n-1$.
 We denote the set of all
admissible tuples by $\scA_{n,k}$.
If the potential $V$ is
integrable, then for each $ [\vd]\in\scD^\star(V)$, the tuple  $\vLambda(\vd)$ is
admissible.  The set of all admissible elements $\scA_{n,k}$ is countable but infinite.

 If
the set of proper Darboux points of a potential $V$ is non-empty, and $N=\card
\scD^\star(V)$, then the $N$-tuples
\begin{equation}
\label{eq:SV}
\scL(V):=\deftuple{\vLambda(\vd)}{[\vd]\in\scD^\star(V)}\in\scC^N_{n-1},
\end{equation}
is called \bfi{the spectrum of} $V$. Let $\scA^N_{n,k}$ be the subset of
$\scC^N_{n-1}$ consisting of $N$-tuples $(\vLambda_1,\ldots,\vLambda_N)$, such that
$\Lambda_i$ is admissible, i.e., $\vLambda_i\in\scA_{n,k}$, for $i=1,\ldots, N$.
We say that the spectrum $\scL(V)$ of a potential $V$ is admissible iff
$\scL(V)\in\scA^N_{n,k}$. The Morales-Ramis Theorem~\ref{thm:MoRa} says that if
potential $V$ is integrable, then its spectrum $\scL(V)$ is admissible.  However, the problem is that
the set of admissible spectra $\scA_{n,k}^N$ is infinite. We showed that from
Theorem~\ref{thm:1} it follows that, in fact, if $V$ is integrable, then its
spectrum $\scL(V)$ belongs to a certain \bfi{finite} subset $\scI^N_{n,k}$ of
$\scA_{n,k}^N$. We call this set  \bfi{distinguished one}, and its elements
\bfi{distinguished spectra}.
In~\cite{mp:08::d} we have proved the following result.
\begin{theorem}
\label{thm:ff}
Let potential $V$ satisfies assumptions of Theorem~\ref{thm:1}. If $V$ is
integrable, then there exists a finite subset $\scI_{n,k}^N\subset \scA_{n,k}^N$, where
$N=\card \scD^\star(V)$, such that $\scL(V)\in \scI_{n,k}^N$.
\end{theorem}
Informally speaking, for fixed $k$ and $n$, we restrict the infinite number of
possibilities in each line of the Morales-Ramis table to a finite set of choices.

\section{Non-generic potentials with $n$ degrees of freedom}
\label{sec:generaln}
If potential $V$ is not generic, then either it possesses an improper Darboux point,  or at least one of its proper Darboux points is not a simple point of $\scD(V)$.    In both cases the Darboux points can be isolated, or they can lie in non-zero dimensional  algebraic subsets in $\CP^{n-1}$.  A description of nongeneric potentials  is simple for $n=2$. For more than two degrees of freedom, even for $k=3$, the problem of classification of nongeneric potentials is very hard and we do not have a general solution of  this problem.

\subsection{Non-square free potentials}
\label{ssec:nsqf}
In~\cite{mp:05::c} it was shown that for $n=2$  a homogeneous potential $V$ is not generic if and only if  it is not square-free.  For $n>2$ the situation is more complicated, but  the implication in one direction is simple. We have the following.
\begin{lemma}
If  a homogeneous potential $V$ is not square-free, i.e., $V=V_0^2V_1$, where $V_0$ is a non-constant homogeneous polynomial, then $V$  is not generic.
\end{lemma}
\begin{proof}
We show that $V$ has improper Darboux points. In fact, we have
\[
 V'(\vq)= V_0(\vq)\left[ 2 V_0'(\vq)V_1(\vq) +V_0(\vq)V_1'(\vq)\right].
\]
All points $[\vd]\in\CP^{n-1}$ such that $V_0(\vd)=0$ are improper Darboux points of $V$. The set of these points is not empty as $\deg V_0>0$.
\end{proof}
We underline  that  the above lemma gives  only sufficient conditions  for the  nongenericity of a potential.  There are examples of potentials without any proper Darboux points (thus `very' nongeneric  ones) which are  square-free. The simplest one is following
\begin{equation}
 V=q_1q_2(q_2-\rmi  q_3).
\label{eq:psqf}
\end{equation}
It has only three improper Darboux points, namely: $[0:0:1]$, $[1:0:0]$ and $[0:\rmi :1]$.

We show that under mild assumptions  non-square free potentials are not integrable.
\begin{theorem}
 \label{thm:ffactor}
Let us consider a homogeneous  potential of the form $V=V_0^2V_1$,   $\deg V=k>2$,  where $V_0$ is a non-constant polynomial. If there exists $\vd\in\C^n$ such that
\begin{equation}
V_0(\vd)=0, \quad V_1(\vd)\neq0, \mtext{and} \sum_{i=1}^n \left( \dfrac{\partial V_0}{\partial q_i}(\vd)\right)^2 \neq
0,
\label{eq:nosquarefree}
\end{equation}
then $V$ is not integrable with rational first integrals.
\end{theorem}
\begin{proof}
Hamilton's equations for potential $V$ have the form
\begin{equation}
 \dot \vq=\vp,\qquad \dot \vp=-V_0(\vq)\left[ 2 V_0'(\vq)V_1(\vq) +V_0(\vq)V_1'(\vq)\right].
\label{eq:hhamy}
\end{equation}
Each  $\vd\in\C^n$  from the following set
\begin{equation}
\label{eq:s}
 \cS(V_0):=\defset{\vd\in\C^n}{V_0(\vd)=0}\setminus\{\vzero\},
\end{equation}
gives a point $[\vd]\in\CP^{n-1}$ which  is  an improper Darboux point of $V$. By Theorem~\ref{thm:im}, if $V$ is integrable, then
for each $\vd\in \cS(V_0)$ matrix $V''(\vd)$ is nilpotent. We show that under assumptions of our theorem it is impossible.

In fact, for $\vd\in\cS(V_0)$  we have
\begin{equation}
 V''(\vd)= 2 V_1(\vd) V'_0(\vd)V'_0(\vd)^T,
\end{equation}
where we understand
\begin{equation*}
 V'(\vq)=\left[\pder{V}{q_1}(\vq), \ldots, \pder{V}{q_n}(\vq)\right]^T.
\end{equation*}
Let $\vd$ satisfies assumptions~\eqref{eq:nosquarefree}. Then $V''(\vd)\neq\vzero_n$.  We show that $V''(\vd)$ is semi-simple and has a non-zero eigenvalue. Let
\begin{equation*}
 \va =V'_0(\vd), \mtext{and} \vA= \va\va^T.
\end{equation*}
As, by assumptions, $\va\neq\vzero$,  and $\va$ is not isotropic, there exist  $\vb_i\in\C^n$ satisfying
\begin{equation*}
  \va^T\vb_i=0, \mtext{for} 1\leq i\leq n-1,
\end{equation*}
and such that $\vb_1, \ldots, \vb_{n-1}, \va$ are $\C$-linearly independent.
We have
\begin{equation}
\vA\vb_i= \va\va^T \vb_i =0, \mtext{for} i=1,\ldots,n-1,
\end{equation}
and
\begin{equation}
\vA\va= \va\va^T\va=( \va^T\va) \va,
\end{equation}
so, $\vb_i$ are eigenvectors of $\vA$ with zero eigenvalue, and  $\va$ is an eigenvector of $\vA$ with eigenvalue $\va^T\va\neq 0$, as we claimed.
\end{proof}
\begin{remark}
 The existence of $\vd$ satisfying~\eqref{eq:nosquarefree}, implies, among other things, that $V_0\nmid V_1$.
\end{remark}
In applications it is important to have a version of Theorem~\ref{thm:ffactor} for cases with prescribed form of factor $V_0$ of the potential. Below we  consider two such cases.
\begin{lemma}
 Let us consider a homogeneous  potential of the form $V=V_0^2V_1$, $\deg V=k>2$,  where
\begin{equation}
 V_0=\sum_{i=1}^n\alpha_iq_i.
\label{eq:linek}
\end{equation}
 If
\begin{equation}
\label{eq:ani}
 \sum_{i=1}^n\alpha_i^2\neq 0,
\end{equation}
and $V_0\nmid V_1$,
then $V$ is not integrable with rational first integrals.
\end{lemma}
\begin{proof}
Thanks to assumption~\eqref{eq:ani}, we can change coordinates in such a way that $V_0= q_n$. So, we consider
potential
\begin{equation}
\label{eq:q_2V}
 V = q_n^2 V_1,
\end{equation}
where $V_1$ is a   homogeneous polynomial not divisible by $q_n$.

For the considered potential, set $\cS(V_0)$  defined by~\eqref{eq:s} consists of all non-zero vectors $\vd$, such that $d_n=0$. We show that  exists $\vd\in\cS(V_0)$ such that $V_1(\vd)\neq 0$.

We can write $V_1$ in the following form
\begin{equation*}
 V_1 = \sum_{s=0}^m W_s q_n^s, \quad m\geq 0,
\end{equation*}
where $W_s$ are homogeneous polynomials in $\widetilde\vq=(q_1,\ldots, q_{n-1})$. By assumption $V_1$ is not divisible by $q_n$, so $W_0\neq 0$.

Assume that for each $\vd\in\cS(V_0)$ we have $V_1(\vd)=0$. This implies that for each  non-zero $\widetilde\vq\in\C^{n-1}$, $W_0(\widetilde\vq)=0$, and so $W_0=0$. This contradiction shows that we have $\vd\in\cS(V_0)$  such that $V_1(\vd)\neq0$. Now, the statement of our lemma  follows from Theorem~\ref{thm:ffactor}.
\end{proof}
\begin{lemma}
 Let us consider a homogeneous  potential of the form $V=V_0^2V_1$, $\deg V=k>2$,  where
\begin{equation}
 V_0=\sum_{i+j=2}^n\alpha_{ij}q_i q_j=\frac{1}{2}\vq^T\vA\vq.
\label{eq:blinek}
\end{equation}
 If $\vA$ is semi-simple, has two non-zero  different eigenvalues,
and $V_0\nmid V_1$,
then $V$ is not integrable with rational first integrals.
\end{lemma}
\begin{proof}
Thanks to assumptions concerning matrix $\vA$ we can choose  coordinates in such a way that
\begin{equation}
 V_0= \sum_{i=1}^n \rho_i q_i^2,
\end{equation}
where $\rho_1, \ldots, \rho_n$ are eigenvalues of $\vA$. We show, by a contradiction,  that for an arbitrary $\vd\in\cS(V_0)$ we have
\begin{equation}
\label{eq:n0}
 \sum_{i=1}\left( \pder{V}{q_i}(\vd)\right)^2\neq 0.
\end{equation}
In fact, if there exists $\vd\in\cS(V_0)$ such that
\begin{equation*}
 \sum_{i=1}\left( \pder{V}{q_i}(\vd)\right)^2= 0,
\end{equation*}
 then its coordinates satisfy
\begin{equation*}
 \begin{split}
  \rho_1d_1^2+\cdots+\rho_nd_n^2 &=0, \\
 \rho_1^2d_1^2+\cdots+\rho_n^2d_n^2 &=0.
 \end{split}
\end{equation*}
Because $\vd\neq\vzero$, this implies that
\begin{equation*}
 \rho_i\rho_j(\rho_i-\rho_j)=0, \mtext{for} 1\leq i,j\leq n.
\end{equation*}
But it is impossible because we assumed that there exist two  non-zero and different eigenvalues of $\vA$. A contradiction proves our claim.

As $V_0\nmid V_1$, there exists $\vd\in\cS(V_0)$, such that $V_1(\vd)\neq0$. Now,  the statement of our lemma follows from Theorem~\ref{thm:ffactor}.
\end{proof}

\subsection{Presence of an improper Darboux point}
Let us assume that potential $V$ possesses an improper Darboux point $[\vd]$, i.e., $V'(\vd)=\vzero$. From the Euler identity it follows that $\vd$ is an eigenvector of the Hessian matrix with the corresponding eigenvalue equal to zero.  The existence of an improper Darboux point gives a restriction on the form of the potential.  At first we describe these restrictions in the most general settings. We perform our  considerations separately for non-isotropic and isotropic Darboux points.
If $\vd$ is non-isotropic, then using a complex rotation we can locate it in $(0,\ldots,0,1)$.
\begin{lemma}
\label{lem:Vimpn}
Let $V$ be  an  homogeneous polynomial potential of degree $k>2$, and $[\vd]=[0:\cdots:1]$ its   improper non-isotropic Darboux point.  Then $V$ has the following form
\begin{equation}
 V =\sum_{i=0}^{k-2}V_{k-i}q_n^i,
\end{equation}
where $V_s$  denotes a homogeneous polynomial of degree $s$ in variables $\widetilde \vq:=(q_1,\ldots,q_{n-1})$.
\end{lemma}
\begin{proof}
 Let us write the potential $V$ in the form
\begin{equation}
 V =\sum_{i=0}^{k}V_{k-i}q_n^i.
\end{equation}
Then
\begin{equation*}
 V'(\vq)= \left( \sum_{i=0}^{k-1}V_{k-i}'(\widetilde \vq)q_n^i, \sum_{i=1}^{k}iV_{k-i}(\widetilde \vq)q_n^{i-1} \right),
\end{equation*}
and thus
\begin{equation*}
 V'(\vd)= \left( V_{1}'(\vzero), kV_{0}( \vzero) \right) =(\vzero, 0).
\end{equation*}
This implies that $V_0=0$ (because $V_0$ is a constant polynomial), and $V_1=0$ (because $V_1$ is a linear form).
\end{proof}
Notice that if we write
\begin{equation}
 V_2(\widetilde\vq)=\frac{1}{2}{\widetilde \vq}^T \vS \widetilde \vq,
\end{equation}
then
\begin{equation}
 V''(\vd)=\begin{bmatrix}
             \vS & \vzero\\
              \vzero^T & 0
          \end{bmatrix}.
\end{equation}
If $\vd$ is an isotropic Darboux point, then using a complex rotation we can locate it in $(0,\ldots,0,\rmi,1)$.
\begin{lemma}
\label{lem:iVimp}
Let $V$ be  an  homogeneous polynomial potential of degree $k>2$.  If $[\vd]$ is its improper isotropic Darboux point  with $\vd=(0,\ldots,0,\rmi,1)$, then $V$ has the following form
\begin{equation}
 V =\sum_{j=0}^{k-2}W_{k-j}(q_n-\rmi q_{n-1})^j,
\end{equation}
where $W_s$  denotes a homogeneous polynomial of degree $s$ in variables $\widetilde \vq:=(q_1,\ldots,q_{n-2}, z_{n-1})$, where  $z_{n-1} :=q_n+\rmi q_{n-1}$.
\end{lemma}
\begin{proof}
 We introduce  new variables $\vq=\vB\vz$, putting
\begin{equation*}
 z_j= q_j, \mtext{for} 1\leq j\leq n-2; \mtext{and} z_{n-1}= q_n+\rmi q_{n-1}, \quad  z_n=q_n-\rmi q_{n-1}.
\end{equation*}
Let $W(\vz):=V(\vB\vz)$, and $\vc:=\vB^{-1}\vd=[0,\ldots,0,2]$. As $W'(\vz)=\vB^T V'(\vB\vz)$,   we have
\begin{equation*}
 W'(\vc) = \vB^T V'(\vB\vc)=\vB^TV'(\vd)=\vzero.
\end{equation*}
Thus, proceeding as in the proof of the previous lemma, we easily  show that $V$ has the prescribed form.
\end{proof}
A further simplification of the form of $V$ with an improper Darboux point can be  done. Namely, using a complex rotation which  fixes the Darboux point we can transform  the quadratic form $V_2$ in  Lemma~\ref{lem:Vimpn} as well as the form $W_2$ in Lemma~\ref{lem:iVimp}, into the normal form. We can do even better. By Theorem~\ref{thm:im}, if $V$ is integrable, then the Hessian matrix $V''(\vd)$ is nilpotent.  Thus, in fact,  the normal forms of   $V_2$ and $W_2$ do not depend on free parameters.
Let us underline that even if all eigenvalues of $V''(\vd)$ vanish,  the fact that it is a symmetric matrix does not imply that  $V''(\vd)$ vanishes.  However, in this particular case if it vanishes  we have the following.
\begin{lemma}
Let $V$ be  an integrable  homogeneous polynomial potential of degree $k=3$. Then it   has an improper  Darboux point $[\vd]$ such that  $V''(\vd)$ vanishes if and only if  it admits a linear first integral $I=\vd^T\vp$.
\end{lemma}
\begin{proof}
 We denote
\begin{equation}
\label{eq:Vi}
 \pder{V}{q_i}=: \frac{1}{2} \vq^T \vF^{i}\vq, \quad {\vF^{i}}^T=\vF^{i},
\end{equation}
for $i=1,\ldots, n$.  Just comparing the mixed derivative of $V$ we easily find that the following identities hold
\begin{equation}
\label{eq:ijl}
 F^i_{jl}=F^j_{il} \mtext{for} 1\leq i,j,l\leq n.
\end{equation}
The Hessian matrix $V''(\vq)$ can be written in the form
\begin{equation}
\label{eq:hessq}
 V''(\vq)= \begin{bmatrix} \vq^T\vF^1 \\ \vdots \\  \vq^T\vF^n
           \end{bmatrix}.
\end{equation}
Assume that $I=\vd^T\vp$ with $\vd\neq \vzero$ is a first integral. Then
\begin{equation}
 \sum_{i=1}^n d_i \pder{V}{q_i}= \frac{1}{2} \vq^T\vB\vq=0,
\end{equation}
where
\begin{equation}
\label{eq:B}
 \vB:= \sum_{i=1}^nd_i\vF^i.
\end{equation}
Thus $\vB=\vzero_n$.  But, using~\eqref{eq:ijl}, we obtain
\begin{equation}
\label{eq:irs}
 0=\sum_{i=1}^nd_i F^i_{rs}=\sum_{i=1}^nd_i F^r_{is}  \mtext{for} 1\leq r,s\leq n,
\end{equation}
that is
\begin{equation}
\label{eq:df}
 \vd^T\vF^r=\vzero, \mtext{for} 1\leq r \leq n.
\end{equation}
The above shows that $V''(\vd)=\vzero_n$, see~\eqref{eq:hessq}. Moreover, \eqref{eq:df} implies that
\begin{equation}
 \vd^T\vF^r\vd = 0 \mtext{for} 1\leq r \leq n,
\end{equation}
and hence $V'(\vd)=\vzero$, see~\eqref{eq:Vi}, so   $[\vd]$ is an improper Darboux point of $V$.

Let us assume that $[\vd]$ is an improper Darboux point of $V$,  and $V''(\vd)=\vzero_n$. Here $\vzero_n$ denotes $n$-dimensional matrix with all entries equal to zero. We have to show that $I=\vd^T\vp$ is a first integral. Notice that $V''(\vd)=\vzero_n$ is equivalent to~\eqref{eq:df}.  But~\eqref{eq:df} is equivalent, by~\eqref{eq:irs}, to $\vB=0$, see~\eqref{eq:B}, and this implies that $I$ is a first integral.
\end{proof}
Notice that the above lemma shows that, in the prescribed situation, we can reduce the problem by one degree of freedom.  We remark that for $k>3$ this lemma is not valid as there are examples of potentials with an improper Darboux point for which $V''(\vd)=\vzero_n$, but $V$  does not have a first integral linear in the momenta.

A very peculiar case appears if the consider potential does not have any proper Darboux point.  For such   potentials  the Morales-Ramis Theorem~\ref{thm:MoRa} is not applicable.   The following two examples  show  that such potentials exist.
\begin{proposition}
Potential
\begin{equation}
 V=(q_2-\rmi q_1)^{2l}(q_1q_2q_3)^l, \mtext{where} l\in\N,
\label{eq:brrr}
 \end{equation}
 does not possess any proper Darboux point.
\end{proposition}
An easy proof of this Proposition we left to the reader.  Let us notice that one can prove, using Theorem~\ref{thm:ffactor},  that the above potential with $l\in\{1,2\}$ is not integrable.

The second example is more interesting as it shows that even in that very specific class of potentials we can find integrable ones.
\begin{proposition}
Potentials
\begin{equation}
 V_{k,l}: =(q_1-\rmi q_2)^l(q_2\pm\rmi q_3)^{k-l},
\label{eq:bezdy}
\end{equation}
with $k>3$,  $l\not\in\{1,k-1, k/2\}$  do not possess any proper Darboux point.
\end{proposition}
A simple proof we left to the reader.

Potential $V_{k,l}$ given by \eqref{eq:bezdy} admits two additional first integrals
\begin{equation}
 I_1=p_1-\rmi p_2\pm p_3,
\label{eq:jedna}
\end{equation}
and
\begin{equation}
\begin{split}
&I_2=(l+1)\rmi p_1^2(q_2\pm \rmi q_3)+p_2^2[-(l+1)q_1\pm(k-l+1)q_3]+(k-l+1)p_3^2(q_1-\rmi q_2)\\
&+p_1p_2[-(l+1)\rmi q_1+(l+1)q_2\pm(k+2)\rmi q_3]+p_1p_3[\pm(l+1)q_1-(k-l+1)q_3]\\
&-\rmi p_2p_3[\pm(k+2)q_1\mp\rmi(k-l+1)q_2-(k-l+1)q_3]\\
&+(k-2l)(q_1-\rmi q_2)^l(q_2\pm\rmi q_3)^{k-l}(q_1-\rmi q_2\pm q_3)
\end{split}
\label{eq:druga}
\end{equation}
However first integrals $H$, $I_1$ and $I_2$ do not commute
\[
 \{I_1,I_2\}=\frac{1}{2}(2l-k)(H-I_1^2),
\]
so  we cannot say that potentials $V_{k,l}$  are integrable.   On the other hand,
it is easy to check that for $l\in\{1, k-1, k, k/2\}\cap\N$ potential $V_{k,l}$ is integrable of even super-integrable.
Moreover, we found that   for $k=7$ and $l=2$, $V_{k,l}$ is integrable. Apart from first integrals $I_1$ \eqref{eq:jedna} and $I_2$ \eqref{eq:druga}, potential $V_{7,2}$ admits one more first integral $I_3$. Here we write the explicit form of $I_3$ for the potential \eqref{eq:bezdy} with the sign $+$ in the second bracket:
\[
\begin{split}
&I_3=\rmi (8 p_3^4 (\rmi q_1 + q_2) +
  2 p_2 (4 p_3^3 (4 q_1 - 3 \rmi q_2 - q_3) +
    12 \rmi p_1 p_3^2 (q_2 + \rmi q_3) \\
&+ 4 p_3 (\rmi q_1 + q_2)
     (q_1 - q_3) (q_2 + \rmi q_3)^6 -
    p_1 (4 q_1 - 3 \rmi q_2 - q_3) (q_2 + \rmi q_3)^7) +
  8 \rmi p_2^4 (q_1 - q_3) \\
&+ 8 p_1 p_3 (q_1 - \rmi q_2)
   (\rmi q_2 - q_3)^7 - 8 p_1 p_3^3 (q_2 + \rmi q_3) -
  4 p_3^2 (q_1 - \rmi q_2)^2 (q_2 + \rmi q_3)^6\\
& +
  (p_1^2 - 2 (\rmi q_1 + q_2)^3 (q_2 + \rmi q_3)^4)
   (q_2 + \rmi q_3)^8 + 8 p_2^3 (p_1 (-\rmi q_2 + q_3) +
    p_3 (-4 q_1 + \rmi q_2 + 3 q_3))\\
& +
  p_2^2 (-24 \rmi p_3^2 (2 q_1 - \rmi q_2 - q_3) +
    24 p_1 p_3 (q_2 + \rmi q_3) + (q_2 + \rmi q_3)^6
     (4 q_1^2 + 3 q_2^2 - 8 q_1 q_3 \\
&+ 6 \rmi q_2 q_3 + q_3^2))).
\end{split}
 \]
The integrability in the Liouville sense is guaranteed by $I_1$ and $I_3$ and we have one more noncommuting first integral $I_2$.
%

\subsection{Potentials  with infinitely many proper Darboux points}
The non-square free potentials considered in Section~\ref{ssec:nsqf} are examples of potentials possessing infinitely many improper Darboux points.   The question is if there are potentials  with infinitely many proper Darboux points. The answer to this question is affirmative.  In \cite{mp:05::d} it was proved that for $n=2$, the only potentials with this property are the radial ones. For $n>2$ the problem of distinguishing potentials with infinitely many proper Darboux points seems to be  very hard.  Below we describe the difficulty of this problem in general settings.

Let us assume that homogeneous potential $V$ of degree $k$ has infinitely many proper Darboux points. Without loss of the generality we can assume that  infinitely many of  them lie  in the chart $(U_1,\theta_1)$.  According to Lemma~\ref{lem:aff},  on this chart  Darboux points  are elements of  the algebraic set $\scA=\scV(g_1,\ldots,g_{n-1})$, where  $g_i$ are polynomials in variables $\widetilde\vx=(x_1,\ldots, x_{n-1})$.
In particular, the family of infinitely many proper Darboux points is a component $C$ of $\scA$ of dimension greather than zero.
 Let $\bI(C)=\langle f_1,\ldots,f_s\rangle$, where $f_1,\ldots,f_s\in\C[\widetilde\vx]$,  be the ideal of $C$. As $g_i$ vanishes  on  common zeros of  $f_1,\ldots,f_s$, by the Hilbert Nullstellensatz Theorem \cite{Cox:97::},  there exist positive integers $m_i$ and polynomials
$b_{ij}\in\C[\widetilde\vx]$ such that
\begin{equation*}
 g_i^{m_i} = \sum_{j=1}^sb_{ij}f_j , \mtext{for} i=1, \ldots, n-1.
\end{equation*}
Unfortunately,  the above considerations are difficult to apply in practice as we do not know explicit forms of polynomials $f_1,\ldots, f_s$, as well as numbers $m_i$ are unknown.  Moreover, from the above equations we want to determine the potential, so we consider them as non-linear partial differential equations for the potential, see definition of $g_i$ in Lemma~\ref{lem:aff}.     Anyway one can find  sufficient conditions for the existence of infinitely many proper Darboux points.
\begin{lemma}
 If a homogeneous potential $V$ is such that polynomials $g_1, \ldots, g_{n-1}$ have a non-constant common factor  $f\in C[\widetilde\vx]$, then $V$ has infinitely many Darboux points.
\end{lemma}
For $n=3$ we have  the following stronger result.
\begin{lemma}
 A  homogeneous potential $V\in\C_k[q_1,q_2,q_3]$  has infinitely many Darboux points if and only if   on one of  the charts $(U_i, \theta_i)$,  polynomials $g_1,g_2\in\C[x_1,x_2]$  have a non-constant common factor.
\end{lemma}
This lemma is a direct consequence of  the well known fact that for $g_1,g_2\in\C[x_1,x_2]$ the set $\scV(g_1,g_2)$ is infinite if and only if $g_1$ and $g_2$ have a nonconstant common factor in $\C[x_1,x_2]$, see e.g. exercise~2 on page 164 in \cite{Cox:97::}, but for greather number of variables the implication is only in one direction.

Thus, we do not known how to effectively characterise general potentials with infinitely many proper Darboux points.
Nevertheless it is worth to consider the following example.
\begin{lemma}
Assume that for a homogeneous potential $V$ conditions $g_1=\cdots =g_s\equiv0$, where $1\leq s\leq n-1$, are satisfied. Then   for $k=2l$,  $V$   is given by
\[
 V=\sum_{i=0}^{l}(q_1^2+\cdots +q_{s+1}^2)^{l-i}V_{2i},
\]
while for $k=2l+1$ it is
\[
 V=\sum_{i=0}^{l}(q_1^2+\cdots +q_{s+1}^2)^{l-i}V_{2i+1},
\]
where $V_{j}$ denotes a homogeneous polynomial of degree $j$ in variables $q_{s+2}, \ldots, q_n$.
\label{lem:gg1}
\end{lemma}
\begin{proof}
We have to solve the following equations
\begin{equation}
 (1+x_1^2)\dfrac{\partial v}{\partial x_i}+x_i\sum_{\substack{j=1\\
j\neq i}}^{n-1} x_j
\dfrac{\partial v}{\partial x_j}-kx_iv=0,\qquad i=1,\ldots,s.
\label{eq:nag1}
\end{equation}
We make the following substitution
\begin{equation*}
 v=v_1(R)v_2(t_2,\ldots,t_{n-1}),
\end{equation*}
where
\begin{equation*}
 R=\sqrt{x_1^2+\cdots+x_s^2+1} \mtext{ and}
 t_j=\frac{x_j}{R}  \mtext{for}  j=s+1,\ldots, n-1.
\end{equation*}
 Since
\[
\begin{split}
 &\dfrac{\partial v}{\partial
x_l}=\dfrac{x_l}{R}\dfrac{\mathrm{d}v_1}{\mathrm{d}r}v_2-\dfrac{x_lv_1}{R^2}
\sum_{j=s+1}^{n-1}t_j\dfrac{\partial v_2}{\partial t_j},\quad\text{for\ } l=1,\ldots,s, \\
& \dfrac{\partial
v}{\partial x_j}=\dfrac{v_1}{R}\dfrac{\partial v_2}{\partial t_j},\quad \text{for\ }
j=s+1,\ldots,n-2,
\end{split}
\]
each equation from the system  \eqref{eq:nag1} reduces to the ordinary differential equation on $v_1$
\[
\dfrac{\mathrm{d}v_1}{v_1(R)}=\dfrac{k}{R}\mathrm{d}R,
\]
with the solution $v_1=\alpha R^k$. Thus the dehomogenisation of the potential takes the form
\begin{equation}
 v=\alpha
(x_1^2+\cdots+x_s^2+1)^{k/2}v_2\left(\dfrac{x_{s+1}}{R},\ldots,\dfrac{x_{n-1}}{R}\right).
\end{equation}
Now we have to force that $v$ is a homogeneous potential of degree $k$. As
result we obtain
\[
 v=
(x_1^2+\cdots+x_s^2+1)^{k/2}\sum_{i_{s+1},\ldots,i_{n-1}}v_{i_{s+1},\ldots,i_{n-1}}\left(\dfrac{x_{s+1}}{
R}\right)^{i_{s+1}}\cdots
\left(\dfrac{x_{n-1}}{R}\right)^{i_{n-1}},\qquad
v_{i_2,\ldots,i_{n-1}}\in\C,
\]
where $i_{s+1},\ldots, i_{n-1}$ are such nonnegative integers that their sum is
\begin{equation}
i_{s+1}+\cdots+ i_{n-1}=\begin{cases}
    2i-1&i=1,\ldots,[k/2] \text{\ \ for\ odd\ } k,\\
  2i&i=0,\ldots,k/2 \text{\ \ for\ even\ } k.
                   \end{cases}
\label{eq:warrr}
\end{equation}
\end{proof}
Potentials from the above lemma   possess  $s(s-1)/2$  first integrals which  are components of the angular momentum
\[
 I_{ij}=q_ip_j-q_jp_i,\qquad 1\leq i<j\leq s+1.
\]
Quantities $I_{ij}$ do  not commute. However,   $I_m=\sum_{j<m}I_{jm}^2$ with $m=2,\ldots,s+1$ form an involutive  set of $s$ first integrals.

From the proof of  Lemma~\ref{lem:gg1} it follows that the potentials given by this lemma  have infinitely many proper Darboux points.
 Moreover those potentials have some peculiar properties. A general discussion of this type of potentials and their integrability properties will be published separately \cite{mp:09::}.
 Here we consider, as example,  the simplest case  $s=1$ and $n=3$. In this case   potential  has the form
\begin{equation}
 V=\sum_{i=0}^{\left[\frac{k}{2}\right]}v_{2i}(q_1^2+q_2^2)^iq_3^{k-2i},\qquad v_{2i}\in\C,
\label{eq:withg1}
\end{equation}
and it admits
 first integral
\[
 I_1=q_1p_2-q_2p_1,
\]
but it is not  necessarily integrable.
With~\eqref{eq:withg1} we associate a two dimensional  homogeneous potential
$\widetilde V\in\C[y_1,y_2]$  of degree $k$ given by
 \begin{equation}
 \widetilde V=\sum_{i=0}^{\left[{k}/{2}\right]}v_{2i}y_1^{2i}y_2^{k-2i},\qquad v_{2i}\in\C.
\label{eq:wtv}
\end{equation}
\begin{lemma}
\label{lem:s1}
 Point $[ \vs]\in\CP^1$,   with  $\vs= ( s_1, s_2)$, $s_1\neq 0$  is a proper Darboux point of  potential~\eqref{eq:wtv} iff and only if  each  point  of  the curve
\begin{equation}
    \Sigma(\vs):= \defset{[ d_1: d_2: d_3] \in \CP^2}{  d_1^2+d_2^2 = s_1^2 \mtext{and } d_3=s_2}\subset\CP^2,
\end{equation}
is a  Darboux points of  potential~\eqref{eq:withg1}.  Moreover, if matrix $\widetilde V''(\vs)$ has eigenvalues $\lambda_1=\lambda_1(\vs)$ and $\lambda_2=k-1$, then  for an arbitrary $[\vd]\in\Sigma(\vs)$, matrix $V''(\vd)$ has eigenvalues $(0, \lambda_1(\vs), k-1)$.
\end{lemma}
\begin{lemma}
\label{lem:s2}
 Point $[ \vs]\in\CP^1$,   with  $\vs= ( 0, s)$, $s\neq 0$  is a proper Darboux point of  potential~\eqref{eq:wtv} if and only if    point
\begin{equation}
    [\vd(s)]:= [ 0 : 0: s] \in \CP^2,
\end{equation}
is a  Darboux points of  potential~\eqref{eq:withg1}.  Moreover, if matrix $\widetilde V''(\vs)$ has eigenvalues $\lambda_1=\lambda_1(\vs)$ and $\lambda_2=k-1$, then matrix $V''(\vd(s))$ has eigenvalues $( \lambda_1(\vs), \lambda_1(\vs), k-1)$.
\end{lemma}
A direct proof of the above two lemmas we left to the reader.

Let us remark that Lemma~\ref{lem:s1} and~Lemma~\ref{lem:s2}   show that  the Morales-Ramis Theorem~\ref{thm:MoRa} gives the same  necessary conditions  for integrability of  potential ~\eqref{eq:wtv} and potential~\eqref{eq:withg1}.  Moreover, although potential ~\eqref{eq:withg1} has  non-isolated Darboux points, there exists  universal relation between  eigenvalues of $V''(\vd)$  taken over of all Darboux points. In fact, assume that potential~\eqref{eq:wtv} has $k$ proper Darboux points $[\vs_i]\in\CP^1$. Then, then by Theorem~1.2 from \cite{mp:05::c},  we have the following relation
\begin{equation}
 \sum_{i=1}^k\dfrac{1}{\Lambda_i}=-1,
\label{eq:relka23}
\end{equation}
where $ \Lambda_i = \lambda_1(\vs_i) -1$ for $i=1, \ldots, k$.

One can suspect that if potential~\eqref{eq:wtv} is integrable,  then  potential \eqref{eq:withg1} is also integrable.    Let us consider several  examples.

For $k=3$ and $n=2$  there are exactly  three  potentials $\widetilde V_1$,  $\widetilde V_2$  and  $\widetilde V_3$ of the form~\eqref{eq:wtv}  which are integrable and have  $k=3$ proper Darboux points. Below we list them  with the respective first integrals
\begin{align}
&\widetilde V_1=y_1^2 y_2 + \dfrac{1}{3}y_2^3, &&
\widetilde I_2=3 z_1 z_2 + y_1^3 + 3 y_1 y_2^2,\\
&\widetilde V_2=y_1^2 y_2 + 2 y_2^3,&& \widetilde I_2=4 z_1 (z_2 y_1 - z_1 y_2)+y_1^4+4 y_1^2 y_2^2, \\
&\widetilde V_3=y_1^2 y_2 + \dfrac{16}{3} y_2^3,&&
\widetilde I_2=9 (4 z_1^4 + 2 z_1^2 z_2^2 + z_2^4 - 4 z_1 z_2 y_1^3) +
 36 (4 z_1^2 + z_2^2) y_1^2 y_2 \nonumber\\
&&&+ 192 (z_1^2 + z_2^2) y_2^3-6 y_1^6
+ 384 y_1^2 y_2^4 + 1024 y_2^6,
\end{align}
where $z_1$ and $z_2$ are the canonical momenta conjugated to $y_1$ and $y_2$.
It appears that potentials  $V_i$ corresponding to  $\widetilde V_i$ by~\eqref{eq:withg1} are also integrable with polynomial first integrals.
These potentials  and the missing first integrals have the form
\begin{align}
\label{v1}
 V_1&=(q_1^2 + q_2^2) q_3 + \dfrac{1}{3}q_3^3, \\
  I_2&=9 (p_1^4 + p_2^4 + p_3^4) - 6 (q_1^2 + q_2^2)^3 -
 30 (q_1^2 + q_2^2) q_3^4 + 4 q_3^6  +
 12 p_3^2 q_3 (3 (q_1^2 + q_2^2) + q_3^2), \notag
\end{align}
\begin{align}
\label{v2}
V_2&=  q_3 (q_1^2 + q_2^2 + 2 q_3^2),\\
I_2&=4 p_1(p_3 q_1 -  p_1 q_3) +
 4 p_2 (p_3 q_2 - p_2 q_3) + (q_1^2 + q_2^2) (q_1^2 + q_2^2 + 4 q_3^2), \notag
\end{align}
\begin{align}
\label{v3}
&V_3=  (q_1^2 + q_2^2) q_3 + \dfrac{16}{3} q_3^3,  \\
&I_2=36 (p_1^4 + p_2^4) - 36 p_2 p_3 q_2 (q_1^2 + q_2^2) - 6 (q_1^2 + q_2^2)^3 -
 36 p_1 q_1 (p_3 (q_1^2 + q_2^2) - 4 p_2 q_2 q_3)  \notag \\
&+ (3 p_3^2 +
    32 q_3^3) (3 p_3^2 + 12 (q_1^2 + q_2^2) q_3 + 32 q_3^3) +
 6 p_1^2 (12 p_2^2 + 3 p_3^2 + 4 q_3 (6 q_1^2 + 3 q_2^2 + 8 q_3^2))\notag \\
&
+
 6 p_2^2 (3 p_3^2 + 4 q_3 (3 q_1^2 + 6 q_2^2 + 8 q_3^2)). \notag
\end{align}
We note that there is  not a direct relation between first integrals $\widetilde I_2$ and the corresponding  ones $I_2$ as they can have different degrees with respect to the momenta.   We observed also this phenomenon for $k$ higher than three.  It justifies  the following conjecture.
\begin{conjecture}
A three dimensional potential ~\eqref{eq:withg1} is integrable if and only if the corresponding two dimensional potential~\eqref{eq:wtv} is integrable
\end{conjecture}
We mention that potentials \eqref{v2} and \eqref{v3} appeared in paper \cite{Grammaticos:85::}.

\section{Integrability of nongeneric three dimensional  homogeneous potentials of degree three}
\label{sec:threeclass}
The general form of a three dimensional  homogeneous potential of  degree three is following
\begin{equation}
 V=a_1q_1^3+a_2q_1^2q_2+a_3q_1^2q_3+a_4q_1q_2^2+a_5q_2^3+a_6q_2^2q_3+a_7q_3^3+a_8q_1q_3^2+a_9q_2q_3^2+a_{10}q_1q_2q_3.
\label{eq:pot3gen}
\end{equation}
In a generic case  the above potential admits seven proper Darboux points and a complete integrability analysis of such potentials in this generic case was performed in \cite{mp:08::d}.
In this section we investigate the  integrability  of nongeneric  cases of  potential~\eqref{eq:pot3gen}. This analysis is much more complicated and difficult than that for the generic situation.

A classification of nongeneric potentials  can be done  in several ways.  However for us  the classification itself is not so important -- our aim is to distinguish all integrable potentials.  Nevertheless,  in order to perform our  futher considerations in more or less systematic way  we need  a certain  rough classification.

 Potential~\eqref{eq:pot3gen} is not generic  if and only if  the number of its isolated proper Darboux points is smaller than seven.
 It occurs in two exclusive  cases: either  all proper Darboux points  of the potential are simple and then it necessarily possesses an improper Darboux point, or   the potential admits a non-simple  proper Darboux point.  In our considerations we always distinguish one Darboux point.
If all proper Darboux points are simple, then we distinguish  an improper Darboux point.  If there is no any improper Darboux point, then we distinguish a non-simple Darboux  point.  If there is a choice, then the distinguished point is chosen to be  non-isotropic one.

Applying the above rules   we consider separately the following cases:
\begin{description}
 \item[Case A.]  All proper Darboux  points of potential $V$ given by~\eqref{eq:pot3gen} are simple and  $V$ has an improper  Darboux point  which is  not  isotropic.
\item[Case B.] All proper Darboux  points of potential $V$ given by~\eqref{eq:pot3gen} are simple; $V$ has an improper  Darboux point  which is isotropic and it does not have any non-isotropic improper Darboux point.
\item[Case C.] Potential $V$ given by~\eqref{eq:pot3gen} admits an isolated   multiple  proper  non-isotropic Darboux point.
\item[Case D.] Potential $V$ given by~\eqref{eq:pot3gen} admits an isolated   multiple  proper  isotropic Darboux point.
\item[Case E.] Potential $V$ given by~\eqref{eq:pot3gen} admits non-isolated proper Darboux point.
\end{description}
Results of our analysis of Case A can be formulated in the following theorem.
\begin{theorem}
\label{thm:caseA}
 Assume that potential $V$ given by~\eqref{eq:pot3gen} satisfies conditions of Case A    and is integrable. Then $V$ admits a first integral which is a  linear form in $\C[\vp]$.
\end{theorem}
In other words, the above theorem says that in Case A the integrability of the potential always reduces to investigation of a two dimensional situation.

More complicated is Case B in which the integrability properties depend on the rank of the Hessian matrix  $V''(\vd_0)$
where  $[\vd_0]$ is the improper isotropic Darboux point of $V$. We summarise results in the following theorem and  two conjectures.
\begin{theorem}
\label{thm:caseB}
 Assume that potential $V$ given by~\eqref{eq:pot3gen} satisfies conditions of Case B and $[\vd_0]$ is its  improper isotropic Darboux point  $\vd_0$, and   $\rank V''(\vd_0)=0$. Then $V$ admits a first integral  which is a  linear form in $\C[\vp]$.
\end{theorem}
We have more interesting situation if  $\rank V''(\vd_0)=1$. In this case we show that the necessary integrability conditions give the following form of the potential
\begin{equation*}
 V_{\lambda}:=\dfrac{1}{4}(q_2 - \rmi q_3) [\lambda q_1^2 + 2 (q_2 - \rmi q_3) (2 a_4 q_1 + 2 a_5 (q_2 - \rmi q_3) + \rmi q_3)].
\end{equation*}
This potential admits the first integral
\begin{equation*}
 I_0=3 (p_2 - \rmi p_3)^2 + (q_2 - \rmi q_3)^3.
\end{equation*}
Parameter $\lambda$ is an eigenvalue of $V_\lambda''(\vd)$, where $[\vd]$ is the only proper Darboux point of this potential. So, the necessary condition for the integrability of this potential is $\lambda\in\scM_3$.  It seems that  this condition is also sufficient.  More precisely, many tests support the following conjecture.
\begin{conjecture}
\label{con:2}
 Potential~\eqref{eq:exceptio} is integrable if and only if $\lambda\in\mathscr{M}_3\setminus\{1\}$. Moreover,
\begin{enumerate}
 \item  if $V_{\lambda}$ is integrable, then it is integrable with polynomial additional first integrals $I_0$ and $I_2$.
\item for an arbitrary $N\in\N$ we find  $\lambda\in\mathscr{M}_3\setminus\{1\}$ such that $V_\lambda$ is integrable with polynomial commuting first integrals  $F_1$, $F_2$ and $F_3$, and  for an arbitrary choice of  these  integrals  we have $\max_i \deg F_i>N$;
\item for \[\lambda\in \mathscr{M}_3\setminus \defset{ p + \dfrac{3}{2}p(p-1)}{ p\in\Z}, \qquad  \lambda\neq1,\] potential $V_\lambda$ admits four functionally independent polynomial first integrals such that three of them pairwise commute.
\end{enumerate}
\end{conjecture}
If $\rank V''(\vd_0)=2$, then  our investigations strongly support the following conjecture.
\begin{conjecture}
 If $V$ satisfies assumptions of Theorem~\ref{thm:caseB} and $\rank V''(\vd_0)=2$, then $V$ is not integrable.
\end{conjecture}

For Case C we can formulate only the following conjecture supported by many tests performed with the help of  higher order variational equations along a particular solution corresponding to the multiple Darboux point.
\begin{conjecture}
If potential $V$  admits an isolated  multiple  proper  non-isotropic Darboux point, then it is non-integrable.
\end{conjecture}

For Case D the integrability analysis is complete.
\begin{theorem}
If potential $V$  admits an isolated  multiple  proper  isotropic Darboux point, then it is non-integrable except for one one-parameter family of integrable potentials
\begin{equation*}
 V=\dfrac{1}{4}(q_1^2 + q_2^2 + q_3^2)[ 4a_8q_1 -\rmi (1 +4a_8^2)q_2 + (1 - 4a_8^2)q_3].
\end{equation*}
\end{theorem}
For Case E our analysis is  complete  and is summarised  in the following theorem.
\begin{theorem}
In class of  potentials   admitting a  non-isolated proper Darboux point  those integrable are equivalent to four  potentials: three axially-symmetric potentials \eqref{v1} -- \eqref{v3},  and one-parameter family
\[
 V=\dfrac{1}{4}(q_2 - \rmi q_3)[q_1^2 + 2 (q_2 - \rmi q_3) (2 a_5 (q_2 - \rmi q_3) + \rmi q_3)].
\]
\end{theorem}

In our analysis we exclude from further investigations potentials which  admit a first integral which is a  linear form in $\C[\vp]$.

\subsection{Case A}
\label{sec:caseA}
In this section we prove Theorem~\ref{thm:caseA}.
For the convenience of the reader   an outline of our  proof is presented  below.

At first we make a kind of  normalisation of the potential. We underline here that   normalisation concerns only potentials which satisfy applicable necessary conditions for the integrability.
\begin{lemma}
\label{lem:nor}
 Assume that potential~\eqref{eq:pot3gen} admits an improper non-isotropic Darboux point and it is integrable.   Then either it is equivalent to the following potential
\begin{equation}
 V=a_1q_1^3+a_2q_1^2q_2+a_4q_1q_2^2+a_5q_2^3+(q_1+\rmi q_2)^2q_3,
\label{eq:zeq}
\end{equation}
or it possesses a first integral which is a  linear form in $\C[\vp]$.
\end{lemma}
Thus, in the remaining part of the proof we analyse potential~\eqref{eq:zeq}.   It  has
an improper Darboux point $[\vd_0]$ with $\vd_0=(0,0,1)$, and,  moreover,  matrix  $V''(\vd_0)$ is nilpotent.  We show this in the proof of Lemma~\ref{lem:nor}. Hence, $[\vd_0]$  is not a simple point of $\scD(V)$. As the number of isolated Darboux points of $V$ is not greater than seven,  the number of proper  Darboux points of~\eqref{eq:zeq} is not greater than five.   The most degenerated  cases  are excluded by the following lemma.
\begin{lemma}
\label{lem:les}
 If  potential~\eqref{eq:zeq} has less than four proper Darboux points and is integrable, then it admits a first integral which is a  linear form in $\C[\vp]$.
\end{lemma}
The most difficult part of  our  considerations concerns cases when $V$ has five or four proper Darboux points.
The crucial step is to show  that among the spectra  $(\Lambda_1^{(i)}, \Lambda_2^{(i)})$  of all proper Darboux points $[\vd^{(i)}]$  of $V$ there exist certain relations.
\begin{lemma}
\label{lem:A5}
If
  potential~\eqref{eq:zeq}  has five simple proper  Darboux points $[\vd_i]\in\CP^2$, then
\begin{equation}
 \left.
 \begin{split}
 &\sum_{i=1}^5\frac{1}{\Lambda_1^{(i)}\Lambda_2^{(i)}}=1,\\
&\sum_{i=1}^5\frac{\Lambda_1^{(i)}+\Lambda_2^{(i)}}{\Lambda_1^{(i)}\Lambda_2^{(i)}}= -4,\\
 &\sum_{i=1}^5\frac{(\Lambda_1^{(i)}+\Lambda_2^{(i)})^2}{\Lambda_1^{(i)}\Lambda_2^{(i)}}=16.
 \end{split}
 \quad\right\}
\label{eq:rele3gen}
\end{equation}
\end{lemma}
\begin{lemma}
\label{lem:A4}
If
  potential~\eqref{eq:zeq}  has four proper  Darboux points $[\vd_i]\in\CP^2$, then
\begin{equation}
 \left.
 \begin{split}
 &\sum_{i=1}^4\frac{1}{\Lambda_1^{(i)}\Lambda_2^{(i)}}=1,\\
&\sum_{i=1}^4\frac{\Lambda_1^{(i)}+\Lambda_2^{(i)}}{\Lambda_1^{(i)}\Lambda_2^{(i)}}= -4,\\
 &\sum_{i=1}^4\frac{(\Lambda_1^{(i)}+\Lambda_2^{(i)})^2}{\Lambda_1^{(i)}\Lambda_2^{(i)}}=8.
 \end{split}
 \quad\right\}
\label{eq:rele3_4D}
\end{equation}
\end{lemma}
To finish the proof we need the following proposition.
\begin{lemma}
\label{lem:45}
 Assume that potential~\eqref{eq:zeq} has five or four simple proper Darboux points. Then it is not integrable.
\end{lemma}

The  proofs of Lemma~\ref{lem:A5} and \ref{lem:A4} are given in the end of Section~\ref{sec:residues}, and the rest of this section contains proofs of the remaining ones.

\subsubsection{Normalisation}
Let us assume that potential~\eqref{eq:pot3gen} has an improper Darboux point $[\vd_0]\in\CP^2$  which is not isotropic.
If additionally $V$ is integrable, then by Theorem~\ref{thm:im}, matrix $V''(\vd_0)$  is  nilpotent, so $\rank V''(\vd_0)\leq 2$. As $\vd_0$ is  not isotropic, we can  assume that $\vd_0=(0,0,1)$, and this implies that
 $a_7=a_8=a_9=0$.    We have
\begin{equation}
 V''(\vd_0)=\begin{bmatrix}
             2a_3&a_{10}&0\\
a_{10}&2a_6&0\\
0&0&0
            \end{bmatrix}.
\label{eq:hessniso}
\end{equation}
Thus, if  $\rank V''(\vd_0)=0$, then $a_3=a_6=a_{10}=0$,  and in this case potential~\eqref{eq:pot3gen} does not depend on $q_3$, so it admits a first integral $I=p_3$.

If $\rank V''(\vd_0)=1 $, then we can assume that $V''(\vd_0) = \vH_1$, see Appendix~\ref{app:A}.  Thus, we have
$ a_{10}=1$ and $a_6=-a_3=\rmi/2$. Hence, up to a simple rescaling $V$ has the form~\eqref{eq:zeq}.

In this way we proved Lemma~\ref{lem:nor}.

\subsubsection{Potentials with first integrals linear in momenta}
\label{sec:liniowce}
Let us assume that potential~\eqref{eq:zeq} admits a first integral of the form
\[
 I=b_1p_1+b_2p_2+b_3p_3,   \mtext{where} \vb=[b_1,b_2,b_3]^T\in\C^3\setminus\{\vzero\}.
\]
Condition
\[
 \Dt I =b_1\dfrac{\partial V}{\partial q_1}+b_2\dfrac{\partial V}{\partial q_2}+b_3\dfrac{\partial V}{\partial q_3} = 0 ,
\]
gives  the following system of linear homogeneous equations
\[
 \vL\vb=\vzero,
\]
where
\[
 \vL=\begin{bmatrix}
 2 \rmi& -2& 0\\
 2& 2 \rmi& 0\\
 a_4& 3 a_5& -1\\
 2 a_2& 2 a_4& 2 \rmi\\
 3 a_1& a_2& 1
 \end{bmatrix}.
\]
It has a non-zero solution $\vb$ iff all third order minors of $\vL$ vanish. This  leads to the following  equations
\begin{equation}
 \begin{split}
&  a_2 + 2 \rmi a_4 - 3 a_5=0,\qquad 3 a_1 + 2 \rmi a_2 - a_4=0,\qquad  3a_1 + \rmi a_2 + a_4 + 3 \rmi a_5=0,\\
&a_2^2 - 3 a_1 a_4 + \rmi a_2 a_4 - a_4^2 - 9 \rmi a_1 a_5 + 3 a_2 a_5=0,
 \end{split}
\label{eq:minorszeq}
\end{equation}
which have a unique solution of the form
\[
 a_1 = \dfrac{1}{3} (a_4-2 \rmi a_2),\qquad a_5 = \dfrac{1}{3} (a_2 + 2 \rmi a_4).
\]
Thus we showed the following.
\begin{proposition}
 Potential~\eqref{eq:zeq} has a first integral which is a  linear form in $\C[\vp]$  iff it is of the form
\begin{equation}
 V=\dfrac{1}{3} (q_1 + \rmi q_2)^2 [(a_4-2 \rmi a_2) q_1 - (a_2 + 2 \rmi a_4) q_2 + 3 q_3].
\label{eq:lin1}
\end{equation}
\end{proposition}
In fact it is easy to show that potential~\eqref{eq:lin1} is super-integrable.
\subsubsection{Number of Darboux points}
\label{sec:nonisotropic}
For a potential $V$ of degree  3  the proper Darboux points of $V$ are in one-to-one correspondence with non-zero solutions of   $V'(\vq)-\vq=\vzero$.  For potential \eqref{eq:zeq}
these equations  have the form
\begin{equation}
 \begin{split}
 & 3 a_1 q_1^2 + 2 a_2 q_1 q_2 + a_4 q_2^2 + 2 q_1 q_3 + 2 \rmi q_2 q_3-q_1=0,\\
&a_2 q_1^2 + 2 a_4 q_1 q_2 + 3 a_5 q_2^2 + 2 \rmi q_1 q_3 -2 q_2 q_3- q_2=0 ,\quad
(q_1 + \rmi q_2)^2 - q_3=0.
 \end{split}
\end{equation}
The last equation $q_3 = (q_1 + \rmi q_2)^2$,  allows to eliminate $q_3$ from the first two, which  can be written in the following form
\begin{equation}
 \begin{split}
 &P_1:=2 q_1^3 + 3 q_1^2 (a_1 + 2 \rmi q_2) + (a_4 - 2 \rmi q_2) q_2^2 +
 q_1 (-1 + 2 (a_2 - 3 q_2) q_2) ,\\
&P_2:=a_2 q_1^2 + 2 \rmi q_1^3 - 6 q_1^2 q_2 + 2 q_1 (a_4 - 3 \rmi q_2) q_2 +
 q_2 (-1 + 3 a_5 q_2 + 2 q_2^2).
 \end{split}
\end{equation}
Each non-zero solution of the above equations gives one proper Darboux point of $V$.  We write
\begin{equation}
\label{eq:Pi}
 P_i = \sum_{j=0}^3p_{i,j}q_2^j, \mtext{where}  p_{i,j}\in\C[q_1],
\end{equation}
and  calculate
the resultant of polynomials $P_1$ and $P_2$ with respect  to variable $q_2$. It is given by
\begin{equation}
\label{eq:resulmain}
  R:=\result (P_1,P_2,q_2)=q_1\sum_{i=0}^5\alpha_iq_1^i,
\end{equation}
where  $\alpha_i$  depend polynomially on coefficients $a_j$ of potential~\eqref{eq:zeq}, and
\begin{equation}
\alpha_0=2 (2 + a_4^2 + 3 \rmi a_4 a_5), \qquad
\alpha_5=-216 (a_1 + \rmi (a_2 + \rmi a_4 - a_5))^3.
\end{equation}
As in expansion~\eqref{eq:Pi} we have $p_{i,3}\in\C^\star$ for $i=1,2$, each root of $R$ gives a solution of $P_1=P_2=0$.
Note that $\deg R=6$ provided that  $\alpha_5\neq 0$.   Hence   we have at least  six solutions of $P_1=P_2=0$.  However,  if  $q_1=0$, then   $P_1=P_2=0$ implies that  $q_2=0$ provided that $\alpha_0\neq 0$.  Hence, generically root $q_1=0$ does not give rise a Darboux point of $V$.

If potential~\eqref{eq:zeq} has five proper Darboux points,  then  $\alpha_5\neq 0$.  In fact, if $\alpha_5=0$, then one can check with a help of computer algebra system that $P_1=P_2=0$ has at most four non-zero solutions.

In the case $\alpha_5=0$ we can express parameter $a_5$  in terms of remaining coefficients of the potential.
In this situation, if $V$ has four proper Darboux points, then
\begin{equation}
 \beta_4:=3 a_1 + 2 \rmi a_2 - a_4\neq 0.
\end{equation}
Again, this fact can be checked with a  help of a computer algebra system.

If $\beta_4=0$, then it is easy to check that the potential possesses at most one proper Darboux point and it has a first integral which is a  linear form in $\C[\vp]$. This last  statement  proves Lemma~\ref{lem:les}

\subsubsection{Proof of Lemma~\ref{lem:45}}
\label{sss}
We  consider second relations in~\eqref{eq:rele3gen} and ~\eqref{eq:rele3_4D}. A direct application of Lemma~3.4 from~\cite{mp:08::d} shows that these relations have at most finite number of admissible solutions.  Next  we can apply algorithm described in Section~4.1 of~\cite{mp:08::d} in order to find all these solutions.  It appears however that  relations~\eqref{eq:rele3gen} and ~\eqref{eq:rele3_4D} do not have  any admissible solution.

\subsection{Case B}
\label{sec:isotropic}
At first we need to perform a normalisation of the potential.
This  is done in the following lemma.
\begin{lemma}
 Assume that potential~\eqref{eq:pot3gen} has an improper isotropic Darboux point and is integrable. Then it is equivalent to one of the listed below
\begin{gather}
 V=a_1 q_1^3 + (q_2 - \rmi q_3)[a_2 q_1^2 + (a_4 q_1 + a_5 (q_2 - \rmi q_3)) (q_2 - \rmi q_3)],
\label{eq:potis1} \\
 V=a_1 q_1^3 +
 \dfrac{1}{2} (q_2 - \rmi q_3) [2 a_2 q_1^2 + (q_2 - \rmi q_3) (2 a_4 q_1 + 2 a_5 (q_2 - \rmi q_3) +
       \rmi q_3)],
\label{eq:potis2} \\
 V=a_1 q_1^3 + (q_2 - \rmi q_3) [a_2 q_1^2 + a_4 q_1 (q_2 - \rmi q_3) +
    a_5 (q_2 - \rmi q_3)^2 + (1 + \rmi) q_1 q_3].
\label{eq:potis3}
\end{gather}
\end{lemma}
\begin{proof}
 Without loss of the generality we can assume that the improper isotropic Darboux point $[\vd_0]\in\CP^2$ is
$[\vd_0]=[0:\rmi:1]$. This  implies  that
\[
 a_{10}=-\rmi(a_4-a_8),\qquad a_9=3a_5-2\rmi a_6,\qquad a_7=-2\rmi a_5-a_6.
\]
Moreover,  matrix $V''(\vd_0)$ has to be nilpotent because by assumption the potential is integrable. This   gives that $a_2=\rmi a_3$.  Now matrix $V''(\vd_0)$ has the following form
\begin{equation}
 V''(\vd_0)=\begin{bmatrix}
0& \rmi (a_4 + a_8)& a_4 + a_8\\
\rmi (a_4 + a_8)& 2 (3 \rmi a_5 + a_6)& 6 a_5 - 2 \rmi a_6\\
a_4 + a_8& 6 a_5 - 2 \rmi a_6& -2 (3 \rmi a_5 + a_6)
          \end{bmatrix}.
\label{eq:hessiso}
\end{equation}
If  $\rank V''(\vd_0)=0$,  then $V''(\vd_0)=\vzero_3$ and $V$ has the form~\eqref{eq:potis1}.  Similarly, if  $\rank V''(\vd_0)=i$, then we can assume that  $V''(\vd_0)=\vH_i$ with $i=1,2$, see Appendix~\ref{app:A}, and this gives us potentials~\eqref{eq:potis2} and~\eqref{eq:potis3}, respectively.
\end{proof}
 We can exclude potential~\eqref{eq:potis1} from our further discussion as it has first integral $I=p_2-\rmi p_3$. Nevertheless,  one can show that this potential  has the following interesting property.
\begin{proposition}
 If $a_1\neq0$, then potential~\eqref{eq:potis1} is not integrable except the case when $a_4= a_2^2/(3a_1) $ and then it is super-integrable.  If $a_1=0$, then potential~\eqref{eq:potis1} admits two first integrals which commute  provided $a_2=0$.
\end{proposition}
\subsubsection{Potential~\eqref{eq:potis2}}
\label{sec:potis2e}
\begin{lemma}
Assume that $a_1\neq 0$.   Then potential \eqref{eq:potis2} is integrable if and only if $a_2 - 2 a_2^2 + 6 a_1 a_4=0$.
\end{lemma}
\begin{proof}
If $a_1\neq 0$, then potential \eqref{eq:potis2} has three proper Darboux points. Among them there is  the following one
\[
 [\vd]:= \left[ \dfrac{1}{3 a_1}: \dfrac{a_2}{9 a_1^2}: -\dfrac{\rmi a_2}{9 a_1^2}\right].
\]
The Hessian  matrix $V''(\vd)$ has eigenvalues $(0,0,2)$,   and is not diagonalisable except  the case $a_2 - 2 a_2^2 + 6 a_1 a_4=0$. Hence, by Remark \ref{rem:nd}, if    $a_2 - 2 a_2^2 + 6 a_1 a_4\neq 0$, then the potential is not integrable.  If $a_2 - 2 a_2^2 + 6 a_1 a_4=0$, then we can eliminate $a_4$, and the  potential reads
\begin{equation}
 V=a_1 q_1^3 + \dfrac{1}{2}(q_2 - \rmi q_3)\left[2 a_2 q_1^2 + (q_2 - \rmi q_3) \left(\dfrac{a_2 (2 a_2-1)}{3 a_1} q_1 +
    2 a_5 (q_2 - \rmi q_3) + \rmi q_3\right)\right],
\end{equation}
and it is  integrable with the following two commuting first integrals
\[
 \begin{split}
I_1&=  (1 - 2 a_2)^2 a_2^2 (p_2 - \rmi p_3)^2-6 a_1 a_2 (-1 + 2 a_2) p_1 (p_2 - \rmi p_3)+18 a_1^3 q_1^3 +
 9 a_1^2 (p_1^2\\
& + 2 a_2 ((1 - 6 a_5) p_2^2
+ 12 \rmi a_5 p_2 p_3 + (1 + 6 a_5) p_3^2 + (q_2 - \rmi q_3) (q_1^2 +
          q_3 (\rmi q_2 + q_3)))),\\
I_2&=3 (p_2 - \rmi p_3)^2 + (q_2 - \rmi q_3)^3.
 \end{split}
\]
\end{proof}

In the case $a_1=0$  we denote $a_2= \lambda/4$, and then potential \eqref{eq:potis2} can be rewritten in the following form
\begin{equation}
 V_{\lambda}:=\dfrac{1}{4}(q_2 - \rmi q_3) [\lambda q_1^2 + 2 (q_2 - \rmi q_3) (2 a_4 q_1 + 2 a_5 (q_2 - \rmi q_3) + \rmi q_3)].
\label{eq:exceptio}
\end{equation}
If $\lambda\neq 1$, then this potential has exactly one proper Darboux point 
\[
 [\vd_1]=\left[-\dfrac{4 a_4}{\lambda-1}:\dfrac{4 ((1 - 3 a_5) (\lambda-1 )^2 + a_4^2 (3\lambda-4 ))}{(\lambda-1)^2}:-\dfrac{2 \rmi ((1 - 6 a_5) (\lambda-1)^2 + a_4^2 (6\lambda-8))}{(\lambda-1)^2}\right].
\]
 At this point matrix $V''_{\lambda}(\vd_1)$ is semi-simple and  has eigenvalues  $(\lambda,2,2)$.
For all values of the parameter   potential~\eqref{eq:exceptio}  has  first integral
\begin{equation}
\label{eq:I0}
 I_0=3 (p_2 - \rmi p_3)^2 + (q_2 - \rmi q_3)^3.
\end{equation}
The  necessary conditions of the Morales-Ramis  Theorem~\ref{thm:MoRa} for the integrability of this potential
have the form
\begin{equation}
\label{eq:lc3}
\begin{split}
\lambda\in \mathscr{M}_3&:=\defset{ p + \dfrac{3}{2}p(p-1)}{ p\in\Z}\cup
\defset{\dfrac 1
{2}\left[\dfrac {2} {3}+3p(p+1)\right]}{p\in\Z}\\
&\cup \defset{-\dfrac 1 {24}+\dfrac 1 {6}\left( 1 +3p\right)^2}{ p\in\Z}
\cup\defset{-\dfrac 1 {24}+\dfrac 3 {32}\left(  1  +4p\right)^2}{ p\in\Z}\\
&\cup\defset{-\dfrac 1 {24}+\dfrac 3 {50}\left(  1  +5p\right)^2}{ p\in\Z}\cup\defset{-\dfrac 1 {24}+\dfrac{3}{50}\left(2 +5p\right)^2}{ p\in\Z}.
\end{split}
\end{equation}
We notice the following amazing fact. In all cases which we were able to check, if the above necessary condition is  satisfied, then in fact the potential  is  integrable or even super-integrable with polynomial first integrals.  Several examples from our experiments are given in Appendix~\ref{app:B}.
Results of these experiments we collected in the Conjecture~\ref{con:2} formulated at the beginning of this section.

If $\lambda=1$,  then potential  \eqref{eq:exceptio} does not have any proper Darboux point provided $a_4\neq0$.
For  $\lambda=1$ and $a_4=0$ potential takes the form \eqref{eq:exceptioinf} and it has infinitely many proper Darboux points and is integrable, see Section~\ref{ssec:caseE}.

\subsubsection{Potential \eqref{eq:potis3}}
\label{sec:iso1}
\paragraph{Three proper Darboux points.}

Potential \eqref{eq:potis3} has  at most three proper Darboux points which lie in the affine part of $\CP^2$.  This fact can be checked  either directly as one can solve explicitly the equations defining proper Darboux points, or  one can show that the multiplicity of the improper Darboux point is at least four.
 If it has three proper Darboux points, then
\begin{equation}
\label{eq:tis3C}
 a_1 ( 3 a_1-1 + \rmi)\neq0,
\end{equation}
 and  among them there is  point $[\vd]\in\CP^2$ with
\begin{equation}
[\vd ]= \left[ \dfrac{1}{3 a_1}: \dfrac{a_2}{ 3 a_1 ( 3 a_1-1 + \rmi)}: \dfrac{a_2}{3 a_1 ( 3 \rmi a_1-1 - \rmi) }\right].
\label{eq:dodo}
\end{equation}
This Darboux point
 has the spectrum
\begin{equation}
  (\Lambda_1,\Lambda_2):=(\Lambda,\Lambda),  \mtext{where} \Lambda=-1 +\dfrac{1 - \rmi}{3 a_1}.
\label{eq:nalambda}
\end{equation}
Note that from the above formula it follows that $\Lambda\neq-1$.
Assuming that $a_1\neq 0$ and using  \eqref{eq:nalambda} we can  rewrite  potential~\eqref{eq:potis3} in the form
\begin{equation}
 V=\dfrac{1 - \rmi}{3(\Lambda+1)} q_1^3 + (q_2 - \rmi q_3) [a_2 q_1^2 + a_4 q_1 (q_2 - \rmi q_3) +
    a_5 (q_2 - \rmi q_3)^2 + (1 + \rmi) q_1 q_3].
\label{eq:potis3rew}
\end{equation}
\begin{lemma}
\label{lem:3r}
 Assume that potential~\eqref{eq:potis3rew} has three simple proper Darboux points $[\vd_1]$,  $[\vd_2]$ and $[\vd_3]:=[\vd]$ with the respective spectra $(\Lambda^{(i)}_1,\Lambda^{(i)}_2)$ for $i=1,2,3$.   Then the  following relations are satisfied
\begin{equation}
\label{eq:3r}
 \left.
 \begin{split}
 &\sum_{i=1}^2\frac{1}{\Lambda_1^{(i)}\Lambda_2^{(i)}}=\dfrac{2 \rmi +2 (1 - \rmi) a_4}{a_2^2},  \\
&\sum_{i=1}^2\frac{\Lambda_1^{(i)}+\Lambda_2^{(i)}}{\Lambda_1^{(i)}\Lambda_2^{(i)}}=-\dfrac{2 [a_2^2 + (2 \rmi + a_2^2 + 2(1 - \rmi) a_4) \Lambda]}{a_2^2 (\Lambda+1)} ,\\
 &\sum_{i=1}^2\frac{(\Lambda_1^{(i)}+\Lambda_2^{(i)})^2}{\Lambda_1^{(i)}\Lambda_2^{(i)}}= \dfrac{8 \Lambda [a_2^2 + (\rmi + a_2^2 + (1 - \rmi) a_4) \Lambda]}{a_2^2 (1 + \Lambda)^2}.
 \end{split}
 \quad\right\}
\end{equation}
\end{lemma}
The above lemma can be proved directly as we can find coordinates of Darboux points explicitly. An alternative proof
based on the multivariable residue calculus is given in section~\ref{sec:lem3r}.

At the first glance it seems that the above relations  are not useful as  their right hand-sides depend on the parameters. Nevertheless, we show the following.
\begin{lemma}
\label{lem:rel3}
 Let potential~\eqref{eq:potis3rew} satisfies assumption of Lemma~\ref{lem:3r}. Then, either
\begin{equation}
\label{eq:neq}
 {
  \Lambda_1^{(1)} \left(2 +  \Lambda_1^{(2)}\right)  \Lambda_2^{(1)} +  \Lambda_1^{(2)}\left (2 +  \Lambda_2^{(1)}\right)  \Lambda_2^{(2)} +  \Lambda_1^{(1)} \left( \Lambda_1^{(2)} +  \Lambda_2^{(1)} + 2  \Lambda_1^{(2)}  \Lambda_2^{(1)}\right)  \Lambda_2^{(2)}}\neq 0,
\end{equation}
and the following relation is satisfied
\begin{equation}
 (\Lambda_1^{(1)} - \Lambda_1^{(2)} + \Lambda_2^{(1)} - \Lambda_2^{(2)})^2 + 4 \Lambda_1^{(1)} \Lambda_1^{(2)} \Lambda_2^{(1)} \Lambda_2^{(2)}=0,
\label{eq:relaca}
\end{equation}
or condition~\eqref{eq:neq} is not satisfied, but then   the relation
\begin{equation}
\label{eq:r-2}
 \dfrac{1}{\Lambda_1^{(1)}}+\dfrac{1}{\Lambda_2^{(1)}}+\dfrac{1}{\Lambda_1^{(2)}}+\dfrac{1}{\Lambda_2^{(2)}}=-2,
\end{equation}
is fulfilled.
\end{lemma}
\begin{proof}
From the first relation~\eqref{eq:3r}  we find that
\[
 a_4=\dfrac{(1+\rmi)\left[a_2^2 \left(\Lambda_1^{(1)}  \Lambda_2^{(1)}+\Lambda_1^{(2)}  \Lambda_2^{(2)} \right) - 2 \rmi \Lambda_1^{(1)}  \Lambda_2^{(1)}\Lambda_1^{(2)}  \Lambda_2^{(2)}\right]}{4\Lambda_1^{(1)}  \Lambda_2^{(1)}\Lambda_1^{(2)}  \Lambda_2^{(2)}}.
\]
We substitute this expression into the remaining relations~\eqref{eq:3r}, and then from the second we calculate $\Lambda$ assuming that inequality~\eqref{eq:neq} holds true.  We obtain the following expression
\begin{equation}
 \Lambda=-\dfrac{ \Lambda_1^{(1)}  \Lambda_1^{(2)}  \Lambda_2^{(1)} +  \Lambda_1^{(2)}  \Lambda_2^{(1)}  \Lambda_2^{(2)} +  \Lambda_1^{(1)} \left( \Lambda_1^{(2)} +  \Lambda_2^{(1)} + 2  \Lambda_1^{(2)}  \Lambda_2^{(1)}\right)  \Lambda_2^{(2)}}{
  \Lambda_1^{(1)} \left(2 +  \Lambda_1^{(2)}\right)  \Lambda_2^{(1)} +  \Lambda_1^{(2)}\left (2 +  \Lambda_2^{(1)}\right)  \Lambda_2^{(2)} +  \Lambda_1^{(1)} \left( \Lambda_1^{(2)} +  \Lambda_2^{(1)} + 2  \Lambda_1^{(2)}  \Lambda_2^{(1)}\right)  \Lambda_2^{(2)}}.
\label{eq:nalambde}
\end{equation}
When we substitute this expression into the last relation then we obtain
\[
 \dfrac{(\Lambda_1^{(1)} - \Lambda_1^{(2)} + \Lambda_2^{(1)} - \Lambda_2^{(2)})^2 + 4 \Lambda_1^{(1)} \Lambda_1^{(2)} \Lambda_2^{(1)} \Lambda_2^{(2)}}{\Lambda_1^{(1)} \Lambda_2^{(1)} +\Lambda_1^{(2)} \Lambda_2^{(2)}}=0,
\]
and this gives  relation~\eqref{eq:relaca}.

On the other hand, if
\begin{equation}
  \Lambda_1^{(1)} \left(2 +  \Lambda_1^{(2)}\right)  \Lambda_2^{(1)} +  \Lambda_1^{(2)}\left (2 +  \Lambda_2^{(1)}\right)  \Lambda_2^{(2)} +  \Lambda_1^{(1)} \left( \Lambda_1^{(2)} +  \Lambda_2^{(1)} + 2  \Lambda_1^{(2)}  \Lambda_2^{(1)}\right)  \Lambda_2^{(2)}=0,
\label{eq:sr1}
\end{equation}
then  necessarily we have also \begin{equation}
  \Lambda_1^{(1)}  \Lambda_1^{(2)}  \Lambda_2^{(1)} +  \Lambda_1^{(2)}  \Lambda_2^{(1)}  \Lambda_2^{(2)} +  \Lambda_1^{(1)} \left( \Lambda_1^{(2)} +  \Lambda_2^{(1)} + 2  \Lambda_1^{(2)}  \Lambda_2^{(1)}\right)  \Lambda_2^{(2)}=0.
\label{eq:sr2}
\end{equation}
Subtracting \eqref{eq:sr1} from \eqref{eq:sr2} we get
\[
  \Lambda_1^{(1)} \Lambda_2^{(1)}+\Lambda_1^{(2)} \Lambda_2^{(2)}=0.
\]
But then both these conditions simplify to
\[
 \Lambda_1^{(2)} \Lambda_2^{(2)}\left[\Lambda_1^{(1)}-\Lambda_1^{(2)}+ \Lambda_2^{(1)}- \Lambda_2^{(2)}-2\Lambda_1^{(2)} \Lambda_2^{(2)}\right]=0.
\]
Since $\Lambda_1^{(2)} \Lambda_2^{(2)}\neq0$ this is equivalent to relation \eqref{eq:r-2}.
\end{proof}
The main result of this section is the following.
\begin{conjecture}
 \label{con:3pd}
Assume that potential~\eqref{eq:potis3rew} has three simple proper Darboux points, then it is not integrable.
\end{conjecture}
A justification of the above conjecture gives almost its full proof.     First of all, the assumption allows us to apply Lemma~\ref{lem:3r}. If the potential is integrable, then either relation~\eqref{eq:relaca}, or ~\eqref{eq:r-2} has an admissible solution. However, relation~\eqref{eq:r-2} does not  have admissible solutions. A proof of this fact follows in the same way as proof of Lemma~\ref{lem:45}, see Section~\ref{sss}.
 Moreover,   relation~\eqref{eq:relaca} have admissible solutions, but all of them have the following form
\begin{equation}
\label{eq:so2}
 \begin{split}
 &\Lambda_1^{(1)}=\pm 1,\quad  \Lambda_2^{(1)}=\mp1,\quad \Lambda_2^{(2)}= \Lambda_1^{(2)}, \mtext{or}\\
&\Lambda_1^{(2)}=\pm 1,\quad  \Lambda_2^{(2)}=\mp1,\quad \Lambda_2^{(1)}= \Lambda_1^{(1)}.
\end{split}
\end{equation}
But then, from \eqref{eq:nalambde}, it follows  that either  $\Lambda=-\Lambda_1^{(1)}$,  or $\Lambda=-\Lambda_1^{(2)}$. This implies that $\Lambda$ as well as $-\Lambda$ is admissible, so either  $ \Lambda=0$, or  $\Lambda=-1$ but then the potential is not well defined and $\Lambda=1$. The first possibility is excluded because it is assumed that all  proper Darboux points are simple.  The second case is also excluded because for it condition~\eqref{eq:neq} is not satisfied.

It remains to show that the only admissible solutions of~\eqref{eq:relaca} are only those of the form given by~\eqref{eq:r-2} or such for that condition~\eqref{eq:neq} is not satisfied.  From the form of~\eqref{eq:relaca} it is clear  that among $\Lambda^{(i)}_j$ either one or three are negative.  There is only finitely many cases that  among $\Lambda^{(i)}_j$ three are negative   and among them  there is no   any admissible one satisfying inequality~\eqref{eq:neq}.  The problem is with cases when  only one among  $\Lambda^{(i)}_j$ is negative.
At first we made a restrictive search of admissible solutions of relation~\eqref{eq:relaca}. Simply we checked   all possibilities taking admissible  $\Lambda^{(i)}_j\in [-1,100] $.  We found only admissible solutions of the form \eqref{eq:so2}.  Later we noticed that in fact one can prove systematically that there is no other solutions but the  proof is very long because we have to consider a lot of separate cases.  The idea is following.
Without any loss of the generality we can assume that
\[
 \Lambda_1^{(1)}\in\left\{-1, -\dfrac{589}{600}, -\dfrac{91}{96}, -\dfrac{7}{8}, -\dfrac{481}{600}, -\dfrac{2}{3}, -\dfrac{301}{600},
-\dfrac{3}{8}, -\dfrac{19}{96}, -\dfrac{49}{600}\right\}.
\]
Then the remaining  three admissible  $\Lambda^{(i)}_j$ are rational numbers of  the form $p$, $p/3$, $p/8$, $p/96$ or $p/600$, where $p$ is a positive  integer.  The problem is that we have to take into account all possibilities for choices of   $\Lambda^{(i)}_j$.   Positive admissible  values  of   $\Lambda^{(i)}_j$   belong to the set that is  the   union of the following disjoint sets
\begin{equation}
 \mathscr{N}:=\bigcup_{i=1}^5 \mathscr{N}_i \subset \Q,
\end{equation}
where
\begin{equation*}
\begin{split}
    \mathscr{N}_1&:=\N, \quad  \mathscr{N}_2:= \defset{p/3}{p\in\N, \quad \gcd(p,3)=1}\quad  \mathscr{N}_3:= \defset{p/8}{p\in\N, \quad \gcd(p,8)=1}, \\
 \mathscr{N}_4&:= \defset{p/96}{p\in\N, \quad \gcd(p,96)=1}, \quad  \mathscr{N}_5:= \defset{p/600}{p\in\N, \quad \gcd(p,600)=1}.
\end{split}
\end{equation*}
 Let us consider  as example the following possibility:  $\Lambda_1^{(1)}=-1$ and  all the remaining $\Lambda^{(i)}_j$ are elements of
$\mathscr{N}_2$, i.e.,
\[
  \Lambda_1^{(1)}=-1,\mtext{and}   \Lambda_2 ^{(1)}=\dfrac{m_1}{3},\qquad  \Lambda_1^{(2)}=\dfrac{m_2}{3},\qquad \Lambda_2^{(2)}=\dfrac{m _3}{3},
\]
where  all positive integers $m_1$, $m_2$ and $m_3$ are relatively prime to $3$.
Then relation  \eqref{eq:relaca} becomes
\[
 3(3+m_1-m_2+m_3)^2=4m_1m_2m_3,
\]
but this means that $3+m_1-m_2+m_3=2m$ for a certain integer $m\in\N$. Then the above equation can be rewritten as
\[
 3m^2=m_1m_2m_3,
\]
and  either  $m_1$,  or $m_2$,  or $m_3$ is  divisible  by 3, but it is a contradiction with assumption that $m_i$ are relatively prime to  $3$.  The other possibilities can be analysed in the similar way however the problem is that we have a lot of such possibilities. The number of choices of three elements among five is $\binom{3+5-1}{3}=\binom{7}{3}=35$ and it gives the number of choices of the memberships of three positive $\Lambda^{(i)}_j$ to five possible sets $\mathscr{N}_j$. This number combined with ten choices of negative  $\Lambda_1^{(1)}$ gives 350 cases to check!

\paragraph{Two proper Darboux points.}
Now, we consider  all the cases in which potential~\eqref{eq:potis3} have two proper and simple Darboux points.
If such a case appears, then either the multiplicity of the improper   Darboux point grows, or  an additional improper  Darboux point appears.
At first we have to distinguish such cases. Applying   resultant analysis for polynomials $g_1$ and $g_2$  defining Darboux points in the affine part of $\CP^2$ one can show the following.
\begin{proposition}
\label{pro:ps32}
If potential~\eqref{eq:potis3} has less then three proper Darboux points, then either $a_1=0$, or
\begin{equation}
a_2 a_5 \Lambda[\rmi + 2(1 - \rmi) a_4 - 2 a_4^2 + 6 a_2 a_5]= 0,
\label{eq:forpotis32}
\end{equation}
where $\Lambda$ is given  in~\eqref{eq:nalambda}.
\end{proposition}
Hence, we have to consider several cases separately.  Result of our analysis is the following.
\begin{lemma}
 \label{lem:pt32}
If potential~\eqref{eq:potis3} is integrable, has two proper and simple Darboux points and its remaining Darboux points are improper, then
\begin{equation}
 \label{eq:wd2}
a_5 [1 - (1 + \rmi) a_4]\neq 0,
\end{equation}
and it has the form
\begin{equation}
 V=\dfrac{1 - \rmi}{3(1 + \Lambda)} q_1^3 + a_5 (q_2 - \rmi q_3)^3 +
 q_1 (q_2 - \rmi q_3)\left [a_4 (q_2 - \rmi q_3) + (1 + \rmi) q_3\right],
\label{eq:twoo15}
\end{equation}
where $ \Lambda\neq -1$.
\end{lemma}
\begin{proof}
 At first we notice that potential~\eqref{eq:twoo15} is just potential~\eqref{eq:potis3rew} with $a_2=0$.   Under imposed restrictions on the remaining parameters it has two proper Darboux points.  To prove our lemma we have to show that in remaining cases  given by Proposition~\ref{pro:ps32} potential ~\eqref{eq:potis3} is either non-integrable,  or it does not have two proper and simple Darboux points.

If $a_1=0$, then the potential  has an additional improper Darboux point $[\vd]$ for which $V''(\vd)$ is not nilpotent, so, by Theorem~\ref{thm:im}, in this case $V$ is not integrable.

If $\Lambda=0$ in \eqref{eq:potis3rew}, or, equivalently $a_1=(1-\rmi)/3$  in \eqref{eq:potis3}, then the potential has two proper Darboux points: $[\vd_1]$  with the spectrum $(\Lambda_1^{(1)},\Lambda_2^{(1)})$, and   $[\vd_2]$ with the spectrum $(\Lambda_1^{(2)},\Lambda_2^{(2)})$.   Of course  $\Lambda_i^{(j)}$ depend on the parameters of the potential but the following relations is satisfied
\begin{equation*}
 \sum_{i=1}^2\frac{\Lambda_1^{(i)}+\Lambda_2^{(i)}}{\Lambda_1^{(i)}\Lambda_2^{(i)}}=-2.
\end{equation*}
However, this relation does not have any admissible solution.
One can prove this  fact in the same way as it was made in the proof of Lemma~\ref{lem:45}, see Section~\ref{sss}.  Hence, the potential is not integrable.

Using similar arguments we show that in all  the remaining cases $V$ is either non-integrable or it does not have two proper and simple Darboux points.
\end{proof}
\begin{conjecture}
 Potential~\eqref{eq:twoo15} is not integrable.
\end{conjecture}
This conjecture is justified by the following facts.   Potential~\eqref{eq:twoo15} has two proper Darboux points: $[\vd_1]$  with the spectrum $(\Lambda,\Lambda)$ and   $[\vd_2]$ with the spectrum $(\Lambda_1,\Lambda_2)$. Moreover, the following relation is satisfied
\begin{equation}
 \Lambda_1+\Lambda_2+\dfrac{2\Lambda}{\Lambda+1}=0.
\end{equation}
Taking first few hundreds of smallest admissible values for  $\Lambda$, $ \Lambda_1$,  and $\Lambda_2$ we did not find any admissible solution of this relation. Thus we conjecture that it has no such solutions at all.

\paragraph{One proper Darboux point.}
Analysis performed in the previous point gave an extra output. Namely, the distinguished potentials with two proper Darboux points have these two proper Darboux point under certain conditions imposed on the coefficients of the potential. Thus, if these conditions are not fullfiled, then the potential has at most one proper Darboux point.  We skip this somewhat lengthy but not so difficult analysis   and give here only  the final result.
\begin{proposition}
 If potential~\eqref{eq:potis3} is integrable and has exactly one proper Darboux point  which is simple, then it is
of the form
\begin{equation}
V=\dfrac{1 -\rmi}{3(\Lambda+1)} q_1^3 +
 a_5 (q_2 - \rmi q_3)^3 + \dfrac{1}{2}(1 - \rmi) q_1 (q_2^2 + q_3^2),
 \label{eq:potis3zjednym}
\end{equation}
where $\Lambda\neq -1$ and $\Lambda a_5\neq 0$.
\end{proposition}
Potential~\eqref{eq:potis3zjednym} has one proper Darboux point at
\[
 [\vd]=\left[\dfrac{1}{2}(1 + \rmi) (1 + \Lambda):0:0\right],
\]
and  the spectrum of $[\vd]$  is  $(\Lambda,\Lambda)$. Moreover, $V''(\vd)$ is semi-simple.  Thus, the only necessary condition for the integrability of this potential is this coming from the Morales-Ramis Theorem~\ref{thm:MoRa}, i.e., $\Lambda+1\in\mathscr{M}_3$.  Fixing $\Lambda$ to an admissible values we can look for a polynomial first integral of the system applying the direct method. We performed several such tests  for different choices of $\Lambda$ but we did not find any integrable example.
\subsection{Case C}
As  in the previous cases at first we perform a normalisation of potentials which  belong to the considered class.
\begin{lemma}
 If potential~\eqref{eq:pot3gen} is integrable and it admits an isolated multiple proper Darboux point which is not isotropic, then it is equivalent  to the following potential
\begin{equation}
V=a_1q_1^3+a_2q_1^2q_2+\dfrac{1}{2}q_1^2q_3+a_4q_1q_2^2+a_5q_2^3+a_6q_2^2q_3+\dfrac{1}{3}q_3^3.
\label{eq:multip}
\end{equation}
\end{lemma}
\begin{proof}
Let $[\vd]$ be a multiple proper Darboux point of  potential~\eqref{eq:pot3gen}. As $[\vd]$ is not isotropic we can assume  without loss of the generality that $\vd=(0,0,1)$. This  implies that $a_8=a_9=0$ and $a_7=1/3$. The Hessian matrix $V''(\vd)$ is of the form
\begin{equation}
 V''(\vd)=\begin{bmatrix}
2a_3&a_{10}&0\\
a_{10}&2a_6&0\\
0&0&2
          \end{bmatrix}.
\label{eq:niediag}
\end{equation}
Point $[\vd]$ is a multiple point iff $V''(\vd)$ has an eigenvalue $\lambda=1$.
Now, we have to consider two cases: either $V''(\vd)$ is semi-simple or not.   If $V''(\vd)$ is semi-simple, then we can assume additionally that the coordinates are chosen in such a way that $V''(\vd)$   is diagonal. This imply that $a_{10}=0$. The assumption that  $[\vd]$ is a multiple Darboux point implies that $V''(\vd)$ has an eigenvalue $\lambda=1$.   Without loss of the generality we can assume that $a_3=1/2$ and this gives us potential~\eqref{eq:multip}.

If $V''(\vd)$ is not semi-simple, then it has eigenvalues $\lambda_1=\lambda_2$ and $\lambda_3=2$. As $V''(\vd)$ has eigenvalues $\lambda=1$, we have  $\lambda_1=\lambda_2=1$.  Hence, we can apply Remark~\ref{rem:nd} and show that in this case $V$ is not integrable. A contradiction shows that this case is impossible.
\end{proof}
A necessary condition for the integrability of potential ~\eqref{eq:multip} is $\lambda= 2a_6\in\mathscr{M}_3$,
see~\eqref{eq:lc3}.
Without doubt these conditions are not sufficient. We investigated this question applying the higher order variational equations.  For formulation of the stronger version of the Morales-Ramis theory based on differential Galois groups of higher order variational equations see \cite{Morales:06::,Morales:99::c} and for application of this method to homogeneous potentials see \cite{mp:04::d,mp:05::c,mp:08::d}.
Taking appropriate values for $a_6$ we can apply this method effectively. It appears that in all checked cases (few tens) the considered potential is not integrable for arbitrary values of remaining parameters.  Of course this approach is hopless if we look for  general results.

\subsection{Case D}
At first we show the following.
\begin{lemma}
 If potential~\eqref{eq:pot3gen} is integrable and admits an isolated multiple proper Darboux point which is  isotropic, then it is equivalent to
\begin{equation}
\begin{split}
& V=a_1 q_1^3 + a_2 q_1^2 q_2 + (a_8+\rmi a_{10}) q_1 q_2^2 +
 a_5 q_2^3 + \left(\dfrac{1}{2} - \rmi a_2\right) q_1^2 q_3
+ a_{10} q_1 q_2 q_3+ \\
& \dfrac{1}{2} (2 \rmi a_5
+ (a_{10} - 2 \rmi a_8)^2) q_3^3
-
 \dfrac{1}{2} (-2 + 6 \rmi a_5 + (a_{10} - 2 \rmi a_8)^2) q_2^2 q_3 +
 a_8 q_1 q_3^2 + (-\rmi - 3 a_5 \\
&+ \rmi (a_{10} - 2 \rmi a_8)^2) q_2 q_3^2 .
\end{split}
\label{eq:nieruszalnynew}
\end{equation}
\end{lemma}
\begin{proof}
 Let $[\vd]$ be a multiple proper Darboux point. Because it is isotropic  we can assume that $\vd=(0,\rmi,1)$,  and this implies that
\[
 a_4 = \rmi a_{10} +a_8, \quad a_6 = 1-2\rmi a_5 -a_7, \quad a_9 = 2\rmi a_7 -a_5 -\rmi.
\]
Then $V''(\vd)$ has eigenvalues  $\lambda_1= 2 \rmi (a_2 - \rmi a_3)$ and $\lambda_2=\lambda_3=2$. Moreover,
by the assumption, $V$ is integrable, thus  $V''(\vd)$ is semi-simple and this implies that
\begin{equation}
 4 (-1 + \rmi a_2 + a_3) (a_5 + \rmi a_7) + \rmi (a_{10} - 2 \rmi a_8)^2=0.
\label{eq:srr}
\end{equation}
Additionally, as $[\vd]$ is a multiple point we have
\[
2\rmi (a_2-\rmi a_3)=1.
\]
All the above conditions give rise potential~\eqref{eq:nieruszalnynew}.
\end{proof}
\begin{lemma}
 Potential \eqref{eq:nieruszalnynew} is not integrable except the case when it is equivalent to
\begin{equation}
\label{eq:mint}
 V=\dfrac{1}{4}(q_1^2 + q_2^2 + q_3^2)[ 4a_8q_1 -\rmi (1 +4a_8^2)q_2 + (1 - 4a_8^2)q_3],
\end{equation}
\end{lemma}
\begin{proof}
 We analyse higher order variational equations along the particular solution related to the multiple proper Darboux point.
The absence of the logarithmic terms in the second order variational equations gives
\begin{equation}
 a_1=\dfrac{\rmi}{2}a_{10} + a_8,\qquad
a_2=\dfrac{(-\rmi a_{10} - 2a_8)^3 - 2a_8}{4(a_{10} - 2\rmi a_8)}.
\label{eq:ww2}
\end{equation}
After substitution these values into potential  the absence of logarithms in solutions of the third order variational equations gives the next obstructions
\[
 a_5 = -\dfrac{\rmi}{4}(1 + 4a_8^2),\qquad a_{10}=0,
\]
and the final form of the potential is given by~\eqref{eq:mint}.
It is an integrable potential because it admits   two commuting first integrals
\[
 \begin{split}
  &I_1=p_3 (q_1 + 4 a_8^2 q_1 - 4 \rmi a_8 q_2) + \rmi p_2 ((4 a_8^2-1) q_1 + 4 a_8 q_3) +
 p_1 (\rmi (1 - 4 a_8^2) q_2 - (1 + 4 a_8^2) q_3),\\
&I_2=3 (p_2 + \rmi p_3)^2 +
 12 a_8^2 (4 p_1^2 +
    5 (2 p_2^2 + 2 p_3^2 - \rmi (q_2 - \rmi q_3) (q_2 + \rmi q_3)^2)) +
 12 a_8 (2 \rmi p_1 (p_2 + \rmi p_3)\\
& + q_1 (q_2 + \rmi q_3)^2) + (\rmi q_2 - q_3)^3 +
 192 a_8^5 q_1 (q_2 - \rmi q_3)^2 - 64 \rmi a_8^6 (q_2 - \rmi q_3)^3 +
 48 a_8^4 ((p_2 - \rmi p_3)^2 \\
&+ 5 (q_2 - \rmi q_3)^2 (-\rmi q_2 + q_3)) +
 32 a_8^3 (3 p_1 (\rmi p_2 + p_3) + q_1 (4 q_1^2 + 9 (q_2^2 + q_3^2))).
 \end{split}
\]
Now we consider the case when the denominator in \eqref{eq:ww2} vanishes, i.e.,   when   $a_{10} = 2 \rmi a_8$ in
\eqref{eq:nieruszalnynew}. Then the absence of logarithmic terms in solutions of the second order variational equations
gives $a_1=a_8=0$. Next, the absence of logarithms in solutions of the third order variational equations forces
$a_2=a_5=-\rmi/4$ and potential becomes
\[
 V= \dfrac{1}{4}(-\rmi q_2 + q_3)(q_1^2 + q_2^2 + q_3^2).
\]
But the above potential is exactly potential~\eqref{eq:mint} taken for $a_8=0$ and is integrable.
\end{proof}

\subsection{Case E}
\label{ssec:caseE}
From the analysis performed in all previous subsections we can deduce that  potentials which are integrable and possess a non-isolated proper Darboux points can appear only in Case B.  Moreover, such potentials do not have more than one proper and isolated Darboux point.

Omitting details, for potential \eqref{eq:potis2} appears a one-parameter family of such potentials.
More detailed, for  $\lambda=1$ and $a_4=0$ potential \eqref{eq:exceptio} takes the form
\begin{equation}
 V=\dfrac{1}{4}(q_2 - \rmi q_3)[q_1^2 + 2 (q_2 - \rmi q_3) (2 a_5 (q_2 - \rmi q_3) + \rmi q_3)].
\label{eq:exceptioinf}
\end{equation}
It has infinitely many proper Darboux points
\[
[\vd]=\left[\pm 2\sqrt{2-12a_5-\rmi q_3}:2+\rmi q_3:q_3\right],
\]
with spectrum $\{0,0\}$.  This potential is really integrable with commuting first integrals
\[
 I_1=p_2 q_1- p_1 q_2 + \rmi( p_1 q_3 -  p_3 q_1),\qquad I_2= 3 (p_2 - \rmi p_3)^2 + (q_2 - \rmi q_3)^3.
\]

For potential \eqref{eq:potis3} we selected two families of potentials of this type. Namely,
the first is equivalent to
\begin{equation}
 V=\dfrac{1}{6} [2 q_1^3 + 3(1 + \rmi) a_5 (q_2 - \rmi q_3)^3 + 3 q_1 (q_2^2 + q_3^2)].
\label{eq:info2}
\end{equation}
This potential has infinitely many proper Darboux points
$
 [\vd]=[1: s : \rmi s]
$
 where $s\in\C$, and it is not integrable for all $a_5\in\C$. In fact,  the  Hessian matrix $V''(\vd)$ has  eigenvalues $(1,1,0)$,  and it is not semi-simple.

The second family is following
\begin{equation}
 V= q_1 \left[2 q_1^2 + 3(\Lambda+1)(q_2^2 + q_3^2)\right].
\label{eq:info1}
\end{equation}
This potential has infinitely many proper Darboux points given by
\[
 [\vd_1]=\left[\dfrac{1}{6(\Lambda+1)}:d_2:d_3\right], \mtext{where} d_2^2+d_3^2=\frac{\Lambda}{(1+\Lambda)^3},
\]
with spectrum $(1,\widehat\Lambda)$ where
\begin{equation}
 \widehat\Lambda=-\dfrac{2 \Lambda}{\Lambda+1}.
\label{eq:relat}
\end{equation}
It has also one isolated proper Darboux point
\[
\qquad [\vd_2]=\left[\dfrac{1}{6}:0:0\right],
\]
with spectrum $(\Lambda,\Lambda)$.

Potential \eqref{eq:info1} possesses also one first integral
\[
 I_1=p_3 q_2 - p_2 q_3.
\]
For $\Lambda\neq 0$ we can rewrite \eqref{eq:relat} as
\begin{equation}
\label{rhat}
 \dfrac{2}{ \widehat\Lambda}+\dfrac{1}{\Lambda}=-1.
\end{equation}
If the potential is integrable, then  $\Lambda$ and  $\widehat\Lambda$ belong to $\scM_3$. This implies that we have only a finite number of  choices of  pairs  $( \widehat\Lambda, \Lambda)\in \scM_3\times \scM_3$. In fact, relation~\eqref{rhat} shows that at least one element of pair  $( \widehat\Lambda, \Lambda)$ is negative, but set $\scM_3$ contains only a finite number of negative elements.  So, we can easily find all admissible pairs. There are only three such pairs and all of them give rise integrable potentials.

The first admissible pair is $(\widehat\Lambda,\Lambda)=(-1,1)$, and the corresponding potential is
\begin{equation*}
 V= 2q_1 [q_1^2 + 3 (q_2^2 + q_3^2)].
\end{equation*}
It is easy to check this potential is equivalent to the integrable potential~\eqref{v1}. Similarly, pairs  $(\widehat\Lambda,\Lambda)=(4,-2/3)$  and $(\widehat\Lambda,\Lambda)=(14,-7/8)$ give potentials equivalent to the integrable potentials~\eqref{v2} and~\eqref{v3}, respectively.

Finally we notice that  the relation \eqref{eq:relat}  has also the solution $\Lambda=\widehat\Lambda=0$ that gives
\begin{equation}
 V=\dfrac{1}{6}(1 - \rmi) q_1 [2 q_1^2 + 3 (q_2^2 + q_3^2)].
\label{infon}
\end{equation}
This potential has infinitely many proper Darboux points
\[
 [\vd]=\left[\dfrac{1}{2}(1+\rmi):q_2:\pm \rmi q_2\right],
\]
with spectrum $\{\Lambda_1,\Lambda_2,\Lambda_3\}=\{0,0,1\}$. Since  $V''(\vd)$ is not semi-simple, thus this potential is non-integrable.

\subsection{Relations between spectra of Darboux points}
\label{sec:residues}
Let us recall that the proof of Theorem~\ref{thm:1} given  in \cite{mp:08::d} is based on a version of  the global multi-dimensional residues theorem.   In Theorem~\ref{thm:1} a generic case is considered and this is why  its proof is simple.   In this section we are going to obtain generalisations of relations~\eqref{eq:rkoj} and~\eqref{eq:rtau} for  nongeneric cases.  The main difficulty is connected with the fact that the most important nongeneric cases are at the same time the most degenerated ones.

At first, we briefly  recall basic facts about the multi-dimensional residues and the
Euler-Jacobi-Kronecker formula. For details the reader is refered to
\cite{Aizenberg:83::,Griffiths:76::,Griffiths:78::,Tsikh:92::,Khimshiashvili:06::}.

Let $f_1, \ldots, f_n\in\C[\vx]$  be polynomials of $n$ variables $x_1, \ldots, x_n$,   and $\vc\in\C^n$ be their
 isolated common zero, i.e., $\vc\in\scV(f_1,\ldots,f_n)$.  We denote $\vf:=(f_1,\ldots, f_n)$, and $\scV(\vf):=\scV(f_1,\ldots,f_n)$.
We consider differential $n$-form
\begin{equation}
\label{eq:oloc}
 \omega:= \frac{p(\vx)}{f_1(\vx)\cdots f_n(\vx)} \,
 \rmd x_1\wedge\cdots\wedge \rmd x_n,
\end{equation}
where $p\in \C[\vx]$. The residue of the form
$\omega$ at $\vx=\vc$  is defined by a multiple integral, see e.g. \cite{Griffiths:78::,Griffiths:76::,Khimshiashvili:06::}, but if $\vc$ is a simple point of $\scV(\vf)$, then
\begin{equation}
\label{eq:0res}
  \res(\omega,\vc)= \frac{p(\vc)}{\det \vf'(\vc)}.
\end{equation}
If $\vc\in\scV(\vf)$ is an isolated but not simple point of $\scV(\vf)$, then we cannot use
formula~\eqref{eq:0res} to calculate the residue of the form $\omega$ at this
point. In such a case we can apply a very nice method developed by Biernat in
\cite{Biernat:89::,Biernat:91::} that reduces the calculation of a
multi-dimensional residue to a one dimensional case. We describe it shortly
below.

Without loss of the generality we can assume that $\vc=\vzero$. Let us consider the following analytic set
\begin{equation}
 \scA:=\defset{\vx\in U}{ f_2(\vx)=\cdots=f_n(\vx)=0},
\end{equation}
where $U\subset\C^n$ is a neighbourhood of the origin. Set $\scA$ is a sum of
irreducible one dimensional components $\scA=\scA_1\cup\cdots\cup \scA_m$.
Let $t\mapsto\vvarphi^{(i)}(t)\in\scA_i$, $ \vvarphi^{(i)}(0)=\vzero$, be an injective
parametrisation of $\scA_i$. Then we define the following forms
\begin{equation}
\label{eq:boi}
 w^{(i)} = \frac{p(\vvarphi^{(i)}(t))}{ \det\vf'(\vvarphi^{(i)}(t))} \frac{f_1'(\vvarphi^{(i)}(t))\cdot \dot\vvarphi^{(i)}(t)}{f_1(\vvarphi^{(i)}(t))}
\rmd t.
\end{equation}
As it was shown in \cite{Biernat:91::} we have
\begin{equation}
\label{eq:bre}
 \res(\omega,\vzero)=\sum_{i=1}^m \res(w^{(i)}, 0).
\end{equation}

It appears that the sum of residues  taken  over all points  of  the set $\scV(f)$, under certain assumptions, vanishes.
An example of  general results of this type is   the classical
Euler-Jacobi-Kronecker formula, see,  e.g.,  \cite{Griffiths:78::} and Theorem~3.6 in \cite{mp:08::d}.
However,  this theorem has   too  strong assumptions  concerning polynomials $f_i$.

In order to formulate more general  result we extend the differential form~\eqref{eq:oloc} into  a differential form $\Omega$ in $\CP^n$. To this end we consider $\omega$ as  the expression of $\Omega$ in the affine chart on $\CP^n$.  In order to express $\Omega$ on other charts we use the
standard coordinate transformations.   Poles of  form $\Omega$  can be located in an arbitrary point in $\CP^n$.
They are points of the projective algebraic set $\scV(\vF):=\scV(F_1,\ldots,F_n)\subset
\CP^n$,
  where $F_i$ are homogenisations of $f_i$ and are given by
\begin{equation}
 F_i(z_0,z_1,\ldots, z_n):=z_0^{\deg f_i}f_i\left(\frac{z_1}{z_0},\ldots, \frac{z_n}{z_0}\right), \mtext{for}i=1,\ldots, n.
\end{equation}

Let $(U_i,\theta_i)$ be a chart on $\CP^n$  and $[ \vp]=[p_0:\cdots:p_n]\in U_i\cap\scV(F_1,\ldots,F_n)$. We can define
the residue of the form $\Omega$ at point $[\vp]$ as
\begin{equation}
 \res(\Omega, [\vp]):=\res(\widetilde\omega , \theta_i(  [ \vp] )),
\end{equation}
 where $\widetilde\omega$ denotes form $\Omega$ expressed in the chart
 $(U_i,\theta_i)$.

The form $\Omega$ is defined by homogeneous polynomials $F_1,\ldots,F_m$ and polynomial
\begin{equation}
 P(z_0,z_1,\ldots, z_n):=z_0^{\deg p}p\left(\frac{z_1}{z_0},\ldots,
 \frac{z_n}{z_0}\right).
\end{equation}
To underline the explicit dependence of $\Omega$ on $F_i$ and $P$ we write
symbolically $\Omega=P/\vF$. The following theorem is a special version of the
global residue theorem.
\begin{theorem}
\label{thm:glo}
 Let $\scV(\vF):=\scV(F_1,\ldots,F_n)$ be a finite set. Then for each polynomial $P$ such that
\begin{equation}
 \deg P\leq \sum_{i=1}^n \deg F_i-(n+1),
\end{equation}
we have
\begin{equation}
 \sum_{[\vs]\in\scV(\vF)} \res(P/\vF, [\vs])=0.
\end{equation}
\end{theorem}
For the proof and the more detailed exposition we refer the reader to \cite{Biernat:91::,Griffiths:78::}.

In order to apply the above theorem we set
\begin{equation}
\label{eq:fi}
 f_i = \pder{V}{x_i}-x_i, \qquad i=1, \ldots, n,
\end{equation}
where $V\in\C[\vx]:=\C[x_1,\ldots,x_n]$ is a homogeneous potential of degree $k>2$.     Here we consider $x_1,\ldots, x_n$ as  affine coordinates of a point with homogeneous coordinates $[q_0:q_1:\cdots:q_n]$ in $\CP^n$, i.e.,
\begin{equation*}
 x_i=\frac{q_i}{q_0}, \qquad i=1, \ldots, n.
\end{equation*}
For calculations of the local residues of the form~\eqref{eq:oloc} with  $\vf:=(f_1, \ldots, f_n)$ given by \eqref{eq:fi} we use the following fact proved  in~\cite{mp:08::d}.
\begin{proposition}
\label{pro:ba}
 Point $\vzero\in\scV(\vf)$ is a simple point and $\vf'(\vzero)=-\vE_n$. Thus we have
\begin{equation}
 \res(\omega,\vzero)=(-1)^n p(\vzero).
\end{equation}
If $\vd\in\scV(\vf)$ and $\vd\neq\vzero$, then
\begin{enumerate}
 \item point $[\vd]\in\CP^{n-1}$ is a proper Darboux point of $V$, i.e.,  $[\vd]\in \scD^\star(V)$,
\item the Jacobi matrix $\vf'(\vd)$ has eigenvalues $\Lambda_1(\vd),
\ldots, \Lambda_{n-1}(\vd),\Lambda_n(\vd)=k-2$,
\item if $\det \vf'(\vd)\neq 0$, then
\begin{equation}
 \res(\omega,\vd)= \frac{p(\vd)}{(k-2)\Lambda_1(\vd)\cdots\Lambda_{n-1}(\vd)},
\end{equation}
\item points  $\vd_j:=\varepsilon^j\vd\in\mathscr{V}(\vf)$,  where $\varepsilon$ is a primitive $(k-2)$-root
of the unity, satisfy  $\vf'(\vd_j)=\vf'(\vd)$, for $j=0, \ldots, k-3$.
\end{enumerate}
\end{proposition}

The elements of matrix $\vf'(\vx)$ are  polynomials of degree $k-2$. Thus, we can take
\begin{equation}
\label{eq:pr}
 p_l(\vx):=(\tr \vf'(\vx) -(k-2))^l, \mtext{with}  l\in\{0,\ldots, n-1\}.
\end{equation}
 For this choice of $p_l(\vx)$ we have
\begin{equation}
 p_l(\vd)= \tau_1(\vLambda(\vd))^l \mtext{for} \vd\in{\scD}^\star(V),
\end{equation}
and
\begin{equation}
 p_l(\vzero)= (-n -(k-2))^l.
\end{equation}
Finally, we notice that the homogenisations $\vF$ of $\vf$ are defined by
\begin{equation}
\label{eq:F_i}
 F_i=\pder{V}{q_i}(\vq) - q_0^{k-2} q_i, \mtext{for} i=1, \ldots, n.
\end{equation}
Thus we have to apply Theorem~\ref{thm:glo} to $n$ differential forms $\Omega_l:=P_l/\vF$,  where  $P_l$ is the homogenisation of polynomial~\eqref{eq:pr} with  $l=0,\ldots, n-1$.

It is easy to observe that if $V$ does not have improper Darboux points, then  all poles of the forms $\Omega_l$ are located in the affine part of $\CP^n$ where $q_0\neq 0$. Moreover, if additionally all proper Darboux points are simple, then the  relations~\eqref{eq:rkoj} and \eqref{eq:rtau} in Theorem~\ref{thm:1} follow easily from Proposition~\ref{pro:ba}.

  Now,  we assume that   $n=k=3$,  and we consider a potential $V$ which has  simple proper Darboux points $[\vd_i]\in\CP^2$  with corresponding spectra $(\Lambda_1^{(i)}, \Lambda_2^{(i)})$, $i=1,\ldots, s<7$.   Moreover, we assume also that $V$ has only one improper Darboux point $[\vs]\in\CP^2$ such that $V''(\vs)$ is nilpotent.
Under these assumptions  it is easy to see that the relations  which we are looking for and which follow from the global residue Theorem~\ref{thm:glo},  have the form
\begin{equation}
\label{eq:relsl}
 \sum_{i=1}^s \frac{\left(\Lambda_1^{(i)}+ \Lambda_2^{(i)}\right)^l}{\Lambda_1^{(i)}\Lambda_2^{(i)}}=(-4)^r - S_l,
\end{equation}
where  $S_l$ is the residue  of the form $\Omega_l$ calculate at $[\widehat \vs]:=[0:s_1:s_2:s_3]\in\CP^3$.
In order to calculate $S_l$ we will apply the Biernat formula \eqref{eq:boi} and \eqref{eq:bre}.

\subsubsection{Proof of Lemmata~\ref{lem:A5} and~\ref{lem:A4} }
Potential $V$ given by~\eqref{eq:zeq} has an improper Darboux point located at $[\vs]=[0:0:1]$.
 Hence,   $[\widehat\vs]:=[0:0:0:1]\in\CP^3$ is an element of $\scV(\vF)$, and thus  it is also a pole of  $\Omega_l$.
We calculate the residue of the forms $\Omega_l$  at $[\widehat\vs]$.
To this end we have to change variables in the form
\begin{equation}
\label{eq:oloc3}
 \omega_l(\vx)= \frac{p_l(\vx)}{f_1(\vx)f_2(\vx) f_3(\vx)} \,
 \rmd x_1\wedge\rmd x_2\wedge \rmd x_3,
\end{equation}
in order to pass to the chart of $\CP^3$ containing point  $[\widehat\vs]$.
Form~\eqref{eq:oloc3} is written on the chart $(U_0,\theta_0)$ where $q_0\neq 0$, so
\[
x_1=\dfrac{q_1}{q_0},\qquad  x_2=\dfrac{q_2}{q_0},\qquad x_3=\dfrac{q_3}{q_0},
\]
are coordinates of point $[q_0:q_1:q_2:q_3]\in\CP^3$. Thus, in order to pass into the chart $(U_3,\theta_3)$ where $q_3\neq  0$, we set
\[
 y_1=\dfrac{q_1}{q_3}=\dfrac{x_1}{x_3},\qquad y_2=\dfrac{q_2}{q_3}=\dfrac{x_2}{x_3},\qquad y_3=\dfrac{q_0}{q_3}=\dfrac{1}{x_3}.
\]
The above defines the desired change of variables $\vx\mapsto\vy$.  In  new variables form~\eqref{eq:oloc3} reads
\begin{equation}
\label{eq:ome1}
 \widetilde\omega_l = - \frac{r_l(\vy)y_3^{2-l}}{h_1(\vy)h_2(\vy) h_3(\vy)}\rmd y_1\wedge \rmd y_2 \wedge \rmd y_3,
\end{equation}
where
\[
 h_i(\vy):=y_3^2f_i\left(\dfrac{y_1}{y_3},\dfrac{y_2}{y_3},\dfrac{1}{y_3}\right),\qquad i=1,2,3,
\]
 and
\begin{equation}
\label{eq:hy1}
 r_l(\vy):=  y_3^{l}p_l\left(\dfrac{y_1}{y_3},\dfrac{y_2}{y_3},\dfrac{1}{y_3}\right).
\end{equation}
Notice that the $\vy$-coordinates of point  $[\widehat\vs]$  are $(0,0,0)=\theta_3([\widehat\vs])$.

For further calculations we need the explicit forms of polynomials $h_i$ and $r_l$. We have
\[
\begin{split}
h_1(\vy)&=3 a_1 y_1^2 + y_2 (2 \rmi + a_4 y_2) + y_1 (2 + 2 a_2 y_2 - y_3),\\
h_2(\vy)&=a_2 y_1^2 + 2 y_1 (\rmi + a_4 y_2) + y_2 (-2 + 3 a_5 y_2 - y_3),\\
 h_3(\vy)&=(y_1 + \rmi y_2)^2 - y_3,
\end{split}
\]
and
\[
r_l(\vy)=[2 (3 a_1 y_1 + a_4 y_1 + a_2 y_2 + 3 a_5 y_2 - 2 y_3)]^l,\qquad l=0,1,2.
\]

In order to use the Biernat formula we choose the analytic set
\begin{equation}
\label{anal}
 \scA:=\defset{\vy\in U}{ h_2(\vy)=h_3(\vy)=0},
\end{equation}
 where $U\subset\C^3$ is a neighbourhood of the origin.
 Since
\begin{equation}
\label{ran}
  \rank \dfrac{\partial (h_2,h_3)}{\partial (y_1,y_2,y_3)}(\vzero)=\rank
\begin{bmatrix}
 2\rmi& -2&0\\
0&0&-1
\end{bmatrix}=2,
\end{equation}
set $ \scA$ consists of only one branch $\vvarphi$ passing through the origin. It can be parametrised in the following way
\[
\C\supset W\ni t\longmapsto \vvarphi(t):=\left(-\rmi t+\scO(t^2), t,-\dfrac{1}{4} (a_2 + 2 \rmi a_4 - 3 a_5)^2 t^4+\scO(t^5)\right)\in U\in\C^3,
\]
where $W$ is a sufficiently small neighbourhood of the origin.
This  parametrisation gives the following expressions
\[
\begin{split}
r_l(\vvarphi(t))&=[2 (-3 \rmi a_1 + a_2 - \rmi a_4 + 3 a_5)t+\scO(t^2)]^l, \\
\det \vh'(\vvarphi(t))&=12 (-\rmi a_1 + a_2 + \rmi a_4 - a_5) t+\scO(t^2),
\end{split}
\]
and
\[
 \dfrac{h_1'( \vvarphi(t))\cdot \dot\vvarphi(t)} {h_1(\vvarphi(t))}=\dfrac{2}{t}+\scO(t^0).
\]
To perform further calculations we have to assume that the potential has five proper and simple Darboux points.
Let us  recall, see Section~\ref{sec:nonisotropic},  that this assumption implies that
\begin{equation}
 \label{al5}
\beta_5:=a_1+\rmi a_2-a_4-\rmi a_5\neq 0.
\end{equation}
 It  guarantees  also that the multiplicity of the only improper Darboux point is two, and expansion of  $\det \vh'(\vvarphi)$ starts with linear term in $t$.

Now, we can write the Biernat formula~\eqref{eq:bre} for the form $\widetilde \omega_l$. As the analytic set $\scA$ has only one component, we have
\begin{equation}
 \res(\widetilde \omega_l,\vzero)=\res(w_l,0),
\end{equation}
where
\begin{equation}
\label{wl}
 w_l := -\frac{\varphi_3(t)^{2-l}r_l(\vvarphi(t))}{ \det\vh'(\vvarphi(t))}\, \frac{h_1'(\vvarphi(t))\cdot \dot\vvarphi(t)}{h_1(\vvarphi(t))}\,
\rmd t,
\end{equation}
for $l\in\{0,1,2\}$. Using the derived expansions we easily find that
\[
 \begin{split}
&  w_0=\left(\dfrac{(a_2 + 2 \rmi a_4 - 3 a_5)^4 t^6}{96 (-\rmi a_1 + a_2 + \rmi a_4 - a_5)}+\scO(t^7)\right)\,\mathrm{d}t,\\
&w_1=\left(-\dfrac{(a_2 + 2 \rmi a_4 - 3 a_5)^2 (3 a_1 + \rmi a_2 + a_4 + 3 \rmi a_5)t^3}{
 12 (a_1 + \rmi (a_2 + \rmi a_4 - a_5))}+\scO(t^4)\right)\,\mathrm{d}t,\\
&w_2=\left(\dfrac{2 (-3 \rmi a_1 + a_2 - \rmi a_4 + 3 a_5)^2}{3 (-\rmi a_1 + a_2 + \rmi a_4 - a_5)}+\scO(t^1)\right)\,\mathrm{d}t.
 \end{split}
\]
As these forms are holomorphic at $t=0$, we have
\begin{equation*}
 0=\res(w_l,0)=\res(\widetilde\omega_l,\vzero)=\res(\Omega_l,[\widehat\vs])=S_l,
\end{equation*}
for   $l\in\{0,1,2\}$. In this way we proved Lemma~\ref{lem:A5}, compare formula~\eqref{eq:relsl} and \eqref{eq:rele3gen}.

Now we consider the case when   the potential has four proper and simple Darboux points. Then  we can set  $a_5 = a_2 + \rmi (a_4-a_1)$.
From Section~\ref{sec:nonisotropic} we know that the requirements of the presence of four Darboux points implies
\begin{equation*}
 \beta_4:=3 a_1 + 2 \rmi a_2 - a_4\neq 0.
\end{equation*}
Now, it is important to notice that under these conditions the potential still has only one improper Darboux point which has now multiplicity three.  This implies that, as in the previous case it is enough to calculate the residues of forms $\widetilde \omega_l$ at $\vy=\vzero$, and as in the previous case we use for this purpose the Biernat formula. The analytic set $\scA$ defined by~\eqref{anal} has again only one branch. This follows from formula~\eqref{ran}. However now it has the following parametrisation
 \[
\C\supset W\ni t\longmapsto \vvarphi(t):=\left(-\rmi t+\scO(t^2), t,\dfrac{1}{4} (-3 a_1 - 2 \rmi a_2 + a_4)^2 t^4+\scO(t^5)\right)\in U\subset\C^3,
\]
which gives rise the following expansions
\[
\begin{split}
  r_l(\vvarphi(t))&=[(-12 \rmi a_1 + 8 a_2 + 4 \rmi a_4) t+\scO(t^2)]^l ,\\
\det \vh'(\vvarphi(t))&=6 (-3 a_1 - 2 \rmi a_2 + a_4)^2 t^2+\scO(t^3),
\end{split}
\]
and
\[
 \dfrac{h_1'( \vvarphi(t))\cdot \dot\vvarphi(t)} {h_1(\vvarphi(t))}=\dfrac{3}{t}+\scO(t^0).
\]
Inserting  the above expansions into~\eqref{wl} we obtain
\[
\begin{split}
& w_0=\left(-\dfrac{(-3a_1 - 2\rmi a_2 + a_4)^2t^5}{32}+\scO(t^6)\right)\,\mathrm{d}t,\\
&w_1=\left(- \dfrac{(-3\rmi a_1 + 2a_2 + \rmi a_4)t^2}{2}+\scO(t^3)\right)\,\mathrm{d}t,\\
&w_2=\left(\dfrac{8}{t}+2 (3 \rmi a_1 + a_2 + 2 \rmi a_4)+\scO(t^1)\right)\,\mathrm{d}t.
\end{split}
\]
Thus, we have
\begin{equation*}
 \res(w_0,0)=\res(w_1,0)=S_1=S_2=0, \mtext{and}\res(w_2,0)=S_2=8.
\end{equation*}
The above expressions  and the formula~\eqref{eq:relsl} prove Lemma~\ref{lem:A4}.

\subsubsection{Proof of Lemma~\ref{lem:3r}}
\label{sec:lem3r}
Potential given by~\eqref{eq:potis3rew} possesses the  isotropic improper Darboux point $[\vs]=[0:\rmi:1]$. Hence, point $[\widehat\vs]=[0:0:\rmi:1]\in\CP^3$ belongs to $\scV(\vF)$ and is a pole of forms $\Omega_l$.
Now, we introduce local coordinates  centred at point $[\widehat\vs]$ putting
\[
 y_1=\dfrac{x_1}{x_3},\qquad y_2=\dfrac{x_2}{x_3}-\rmi,\qquad  y_3=\dfrac{1}{x_3}.
\]
Using these coordinates we obtain
\[
 \begin{split}
&h_1=\dfrac{1}{\Lambda+1}[(1 - \rmi) y_1^2 + (1 + \Lambda) y_2 (1 + \rmi + a_4 y_2) + (1 + \Lambda) y_1 (2 a_2 y_2 -
    y_3)], \\
&h_2=a_2 y_1^2 + 3 a_5 y_2^2 + y_1 (1 + \rmi + 2 a)4 y_2) - (\rmi + y_2) y_3,\\
&h_3= -\rmi [a_2 y_1^2 + 3 a_5 y_2^2 + y_1 (1 + \rmi + (-1 + \rmi + 2 a_4) y_2) - \rmi y_3].
 \end{split}
\]
Now we assume that potential \eqref{eq:potis3rew} possesses  three simple proper Darboux points, i.e. $a_2a_5\Lambda[4 (-1 + \rmi + a_4) a_4 - 2 (\rmi + 6 a_2 a_5)]\neq 0$.
Under these assumptions  $[\vs]$ is the only improper Darboux point of the potential, and the analytic set
\begin{equation}
 \scA:=\defset{\vy\in U}{ h_2(\vy)=h_3(\vy)=0},
\end{equation}
has three branches passing through the origin.  Their parametrisations $\vvarphi^{(i)}(t)$ are following
\[
\begin{split}
\C\supset W\ni t\longmapsto \vvarphi^{(1)}(t):=&\left( t,0,(1 - \rmi) t - \rmi a_2 t^2\right)\in U, \\
\C\supset W\ni t\longmapsto \vvarphi^{(2)}(t):=&\left(t, \dfrac{2 a_2 t}{1 - \rmi - 2 a_4 + b},(1 - \rmi) t\right)\in U,\\
\C\supset W\ni t\longmapsto \vvarphi^{(3)}(t):=&\left(t,  \dfrac{2 a_2 t}{1 - \rmi - 2 a_4 - b},(1 - \rmi) t\right)\in U,
\end{split}
\]
where
\[
 b=\sqrt{4 (\rmi-1+ a_4) a_4 - 2 (\rmi + 6 a_2 a_5)}.
\]
Let $w^{(i)}_l$ denotes the Biernat form~\eqref{wl} calculated  for parametrisation $\vvarphi^{(i)}(t)$.  A direct calculations give the following formulae

\begin{align*}
& w_0^{(1)}=\left( \dfrac{(1 + \rmi) (1 + \Lambda)}{a_2 \Lambda t^2} + \dfrac{(\Lambda-1) (\Lambda+1)}{\Lambda^2 t}+\scO(t^0)\right)
\,\mathrm{d}t,\\
&w_0^{(2)}=\left(\dfrac{(1 - \rmi) (1 - \rmi - 2 a_4 + b)^2}{4a_2^2 bt}+\scO(t^0)\right)\,\mathrm{d}t,\\
& w_0^{(3)}=\left(-\dfrac{(1 - \rmi) (-1 + \rmi + 2 a_4 + b)^2}{4a_2^2 bt}
+\scO(t^0)
\right)\,\mathrm{d}t,\\
&  w_1^{(1)}=\left(-\dfrac{2(1 + \rmi)}{a_2 t^2} - \dfrac{2 (1 + 2 \Lambda)}{\Lambda t}+\scO(t^0)\right)\,\mathrm{d}t,\\
& w_1^{(2)}=\left(-\dfrac{(1 - \rmi) (-1 + \rmi + 2 a_4 - b)((1 + \rmi) a_2^2 (1 + \Lambda) +
    \Lambda (-1 + \rmi + 2 a_4 - b)) }{2a_2^2 (1 + \Lambda) bt}+\scO(t^0)\right)\,\mathrm{d}t, \\
& w_1^{(3)}=\left(\dfrac{(1 - \rmi) (-1 + \rmi + 2 a_4 + b) ((1 + \rmi) a_2^2 (1 + \Lambda) +
   \Lambda (-1 + \rmi + 2 a_4 + b))}{2a_2^2 (1 + \Lambda) bt}
+\scO(t^0\right)\,\mathrm{d}t ,
 \end{align*}
\begin{align*}
& w_2^{(1)}=\left(\dfrac{4(1 + \rmi) \Lambda}{a_2 (1 + \Lambda) t^2} + \dfrac{12}{t}+\scO(t^0)\right)\,\mathrm{d}t,\\
& w_2^{(2)}=\dfrac{(1 - \rmi) ((1 + \rmi) a_2^2 (1 + \Lambda) +
   \Lambda (-1 + \rmi + 2 a_4 - b))^2}{a_2^2 (1 + \Lambda)^2 bt}
+\scO(t^0))
\,\mathrm{d}t,\\
& w_2^{(3)}=\left(-\dfrac{(1 - \rmi) ((1 + \rmi) a_2^2 (1 + \Lambda) + \Lambda (-1 + \rmi + 2 a_4 + b))^2}{
 a_2^2 (1 + \Lambda)^2 bt}+\scO(t^0)
\right)
\,\mathrm{d}t.
 \end{align*}
Hence, we have
 \[
\begin{split}
 & S_0=\sum_{i=1}^3 \res(w_0^{(i)},0)=1 + \dfrac{2 \rmi (-1 + (1 + \rmi) a_4)}{a_2^2} - \dfrac{1}{\Lambda^2},\\
 &S_1=\sum_{i=1}^3 \res(w_1^{(i)},0)=-\dfrac{2 [2 \rmi (-1 + (1 + \rmi) a_4) \Lambda^2 + a_2^2 (1 + \Lambda)^2]}{a_2^2 \Lambda (1 + \Lambda)},\\
&S_2=\sum_{i=1}^3 \res(w_2^{(i)},0)=\dfrac{4 [2 \rmi (-1 + (1 + \rmi) a_4) \Lambda^2 + a_2^2 (1 + \Lambda) (3 + \Lambda)]}{a_2^2 (1 + \Lambda)^2}.
\end{split}
 \]
Now,  using formula~\eqref{eq:relsl} with $s=3$ and taking into account that $\Lambda_1^{(3)}=\Lambda_2^{(3)}=\Lambda$, we obtain \eqref{eq:3r}. This finishes the proof.

\section{Final remarks}
Let us compare the  analysis given in this paper with those concerning generic potentials presented in \cite{mp:08::d}.
In the generic case the procedure is well determined for any chosen $n$ and $k$.
The starting points are universal relations between the spectra of Darboux points  of the potential, see Theorem~\ref{thm:1}. They   guarantee the finiteness of the distinguished spectra. Obviously, it is  a very   hard job  to find  all of them but we know that their number is finite. If we already calculated the distinguished spectra,  then we reconstruct the corresponding potentials. Obviously this is also highly non-trivial task  because we have to solve systems of nonlinear equations. For $n=k=3$ each distinguished spectrum gave one integrable potential.

In the nongeneric case there is no such obvious one procedure. The reason is that nongeneric cases have various origins: the multiple proper Darboux points and improper Darboux points.  Additionally the multiplicity of proper as well as improper Darboux points can change in some limits.  This causes that there is no one universal set of relations. The relations exist and we found how to obtain them for the case $n=k=3$ using the multivariable residue calculus.
However,  for bigger $n$ and $k$, an application of the multivariable residue calculus meets highly non-trivial problems. Furthermore,  in some cases the obtained relations depend on the potential coefficients. Thus, they  lost the universal character and we cannot  prove  the finiteness of the distinguished spectra. In some cases, see section~\ref{sec:iso1}, from the set of relations containing  coefficients of potential, see e.g. \eqref{eq:3r}, one can obtain an  universal relation, see e.g. \eqref{eq:relaca}. But its form depends on the considered case and it is unclear if  it gives rise the finiteness of choices of distinguished spectra, see the analysis in the end of section~\ref{sec:iso1}.

It seems that the most difficult for the integrability analysis are potentials with  multiple nonisotropic proper Darboux points, see potential \eqref{eq:multip}. For such potentials one of eigenvalues $\lambda$ of the Hessian matrix $V''(\vd)$ at the multiple Darboux point $[\vd]$  depends on the coefficients of the potential, e.g. for \eqref{eq:multip} we have $\lambda=2a_6$.
This fact causes strong difficulties for the integrability analysis made by means of higher order variational equations along the particular solution defined by this multiple Darboux point. We have the obstruction that $\lambda\in\scM_3$ and obviously, for a given $\lambda\in\scM_3$ higher order variational equations give very quickly a definite answer but the problem is that the set  $\scM_3$ is not finite.  The alternative method is to try to repeat the analysis similar to that performed  for potentials with improper Darboux points. In the most generic case we have six different proper Darboux points and three relations depending on parameters. From them we can deduce    one universal relation but  generally there is no way  to obtain such a universal relation when the number of different proper Darboux points is smaller than six.  The complete analysis will be presented in a separate paper but we only mention that no integrable potential with a multiple isolated proper nonisotropic Darboux point was found.

Concluding, it seems that to  make a complete integrability analysis in nongeneric cases some additional tools and theoretical facts are necessary.

\section*{Acknowledgements}
The author is very grateful to Andrzej J. Maciejewski for many helpful comments
and suggestions concerning improvements and simplifications of some results, as well as for his help in
numerical calculations of distinguished spectra.

The author wishes to thank collegues from Division of Mathematical Physics of N.~Copernicus University, especially Prof. Dr hab. Dariusz Chru\'sci\'nski and Dr Jacek Jurkowski, for their hospitality and the possibility to make some successful calculations of distinguished spectra on their computers.

Special thanks for Alain Albouy, Jean-Pierre Marco, Jacques-Arthur Weil  and Alexei Tsygvintsev for their stimulative questions about the role of improper Darboux points.
The author wishes also to thank Arkadiusz P{\l}oski for his useful comments about multidimensional residues
 and, in particular, for pointing our  attention to papers of G.~Biernat.

This research has been partially supported  by grant No. N N202 2126 33 of Ministry of Science and Higher Education of Poland, by projet de l'Agence National de la
Recherche
``~Int\'egrabilit\'e r\'eelle et complexe en m\'ecanique hamiltonienne''
N$^\circ$~JC05$_-$41465, and  by  UMK grant  414-A.

\appendix
\section{Normal forms of symmetric matrices}
\label{app:A}
In  Chapter~XI of the book~\cite{Gantmaher:88::} the following theorem was proved.
\begin{theorem}
Let  $\vA\in\M(n,\C)$ and $\vA^T=\vA$. There exists $\vU\in\mathrm{O}(n,\C)$ such that   $\vA=\vU \widetilde \vA \vU^{-1}$, where $\widetilde \vA$  is a block-diagonal matrix of the form
\[
 \widetilde \vA=\diag\{\lambda_1\vE_{n_1}+\vS_{n_1},\lambda_2\vE_{n_2}+\vS_{n_2},\ldots,\lambda_p\vE_{n_p}+\vS_{n_p}\},
\]
where $\lambda_1,\lambda_2,\ldots,\lambda_p$ are eigenvalues of $\vS$, $\vE_{n_i}$ denotes the identity matrix of dimension $n_i$, and $\vS_{n_i}$ is a symmetric matrix of the following form
\begin{equation}
 2\vS_{n_i}=\begin{bmatrix}
             0&1&0&\cdots&\cdots&0\\
1&\ddots&\ddots&\ddots&&\vdots\\
0&\ddots&\ddots&\ddots&\ddots&\vdots\\
\vdots&\ddots&\ddots&\ddots&\ddots&0\\
\vdots&&\ddots&\ddots&\ddots&1\\
0&\cdots&\cdots&0&1&0
            \end{bmatrix}+\rmi
\begin{bmatrix}
 0&\cdots&\cdots&0&-1&0\\
\vdots&&\adots&\adots&\adots&1\\
\vdots&\adots&\adots&\adots&\adots&0\\
0&\adots&\adots&\adots&\adots&\vdots\\
-1&\adots&\adots&\adots&&\vdots\\
 0&1&0&\cdots&\cdots&0
\end{bmatrix},
\label{eq:reformy}
\end{equation}
and $\sum_{i=1}^pn_i=n$.
 \label{thm:gant}
\end{theorem}
For $n=3$ this theorem gives the following three possible normal forms of a symmetric nilpotent matrix $\vN\in\M(3,\C)$
\[
\widetilde  \vN_0=\begin{bmatrix}
      0&0&0\\
0&0&0\\
0&0&0
     \end{bmatrix}, \qquad
 \widetilde  \vN_1=\begin{bmatrix}
      -\rmi &1&0\\
\phantom{-}1&\rmi&0\\
0&0&0
     \end{bmatrix},\qquad
\widetilde  \vN_2=\begin{bmatrix}
     0&1-\rmi&0\\
1-\rmi&0&1+\rmi\\
0&1+\rmi&0
    \end{bmatrix}.
\]
However, we need slightly modified normal forms.  Namely we take three matrices $\vP_i\in\mathrm{O}(3,\C)$ where $\vP_0=\vE_3$, and
\[
 \vP_1=\begin{bmatrix}
 0& 0& 1\\
 0& 1& 0\\
 1& 0&0
       \end{bmatrix},\mtext{and}\vP_2=\begin{bmatrix}
       0&0&1\\
1&0&0\\
0&1&0
      \end{bmatrix},
\]
and we set
\begin{equation*}
 \vH_i := \vP_i^T \widetilde\vN_i\vP_i \mtext{for} i=0,1,2.
\end{equation*}
Then we obtain
\[
\vH_0=\begin{bmatrix}
      0&0&0\\
0&0&0\\
0&0&0
     \end{bmatrix}, \qquad
 \vH_1=\begin{bmatrix}
 0& 0& 0\\
 0& \rmi& 1\\
 0& 1& -\rmi
\end{bmatrix},\qquad
\vH_2=\begin{bmatrix}
 0&1+\rmi&1-\rmi\\
1+\rmi&0&0\\
1-\rmi&0&0
\end{bmatrix}.
\]
These new normal forms have the following property
\begin{equation*}
 \vH_i\vd_0=\vd_0  \mtext{for} i =0,1,2,
\end{equation*}
 where $\vd_0=(0,\rmi,1)^T$.

We apply the above facts in the proofs of the following two lemmas.
\begin{lemma}
\label{lem:ss1}
 Let $V\in\C[\vq]=\C[q_1,q_2,q_3]$ be a homogeneous potential of degree $k>2$ having an improper Darboux point $[\vd]\in\CP^2$ with $\vd=(0,0,1)$ such that the Hessian matrix $\vN:=V''(\vd)$ is nilpotent. Then there exists   potential
$\widetilde V$ equivalent to $V$, such that  $[\vd]$ is an improper Darboux point of $\widetilde V$, and  $\widetilde V''(\vd)=\widetilde\vN_i$, where $i=\rank \widetilde V''(\vd)$.
\end{lemma}
\begin{proof}
 Let  $\vN:=V''(\vd)$. At first we notice that $\vd$ is an eigenvector of $\vN$, and thus $\vN$ has the form
\[
 \vN=\begin{bmatrix}
              \vC&\vzero\\
             \vzero^T &0
             \end{bmatrix},
\]
where $\vC$ is a $2\times 2$ nilpotent symmetric matrix, and $\vzero=[0,0]$. Hence $\rank \vN<2$. If $\rank \vN=0$, then $\vN$ is already in the normal form, so we take $\widetilde V=V$ and the lemma is proved. If $\rank \vN=1$, then the matrix  $\vU\in\mathrm{O}(3,\C)$ is of the form
\[
 \vU=\begin{bmatrix}
              \vW&\vzero\\
      \vzero^T & 1
             \end{bmatrix},
\]
 where $\vW\in\mathrm{O}(2,\C)$ is  a matrix which transforms $\vC$ to its normal form.  Notice that $\vU \vd = \vd$.  Now, potential  $\widetilde V(\vq):=V(\vU\vq)$ satisfies the requirement of the lemma and this ends the proof.
\end{proof}
\begin{lemma}
\label{lem:sss1}
 Let $V\in\C[\vq]=\C[q_1,q_2,q_3]$ be a homogeneous potential of degree $k>2$ having an improper Darboux point $[\vd_0]\in\CP^2$ with $\vd_0=(0,\rmi,1)$ such that the Hessian matrix $\vN:=V''(\vd)$ is nilpotent. Then there exists   potential
$\widetilde V$ equivalent to $V$, such that    $[\vd]$ is an improper Darboux point of $\widetilde V$, and  $\widetilde V''(\vd)=\widetilde\vN_i$, where $i=\rank \widetilde V''(\vd)$.
\end{lemma}
\begin{proof}
Obviously the lemma is true when   $\rank \vN=0$. If  $\rank \vN=i>0$, then we take matrix $\vU_i$ which transforms $\vN$ into the normal form $\vH_i$, and set $\widetilde V(\vq):=V(\vU_i\vq)$. One can show that   $\vU_i\vd_0=\alpha_i \vd_0$,  for a certain
$\alpha_i\in\C^\star$,   so
\begin{equation*}
 \widetilde V'(\vd_0)=\vU_i^TV'(\alpha_i\vd_0)=\alpha_i^2\vU_i^TV'(\vd_0)=0.
\end{equation*}
This ends the proof.
\end{proof}

\section{Explicit forms of first integrals for potential \eqref{eq:exceptio}}
\label{app:B}
In section~\ref{sec:potis2e} we formulated the conjecture about the integrability of  potential
\begin{equation}
 V_{\lambda}:=\dfrac{1}{4}(q_2 - \rmi q_3) [\lambda q_1^2 + 2 (q_2 - \rmi q_3) (2 a_4 q_1 + 2 a_5 (q_2 - \rmi q_3) + \rmi q_3)],
\end{equation}
for all $\lambda\in\mathscr{M}_3\setminus\{1\}$, see \eqref{eq:lc3}.
We recall   that it   has one  first integral for all values of the parameters
\begin{equation}
 I_0=3 (p_2 - \rmi p_3)^2 + (q_2 - \rmi q_3)^3,
\end{equation}
and we show some examples of supplementary first integrals.
If $\lambda$ is given by the first item of the Morales-Ramis table~\eqref{eq:tabMoRa} i.e.
\[
 \lambda=p + \dfrac{3}{2}p(p - 1),\qquad p\in\Z,
\]
then for $p=0$ potential  is super-integrable with  commuting first integrals $\{I_0,I_1\}=0$, and for $|p|\geq 1$ is only integrable. Explicit forms for small $p$ are the following.
\begin{itemize}
\item   $p=0$,  $\lambda=0$
\[
 \begin{split}
&\hspace*{-1cm}I_1=p_1 - 2 a_4 p_2 + 2 \rmi a_4 p_3,\\
&\hspace*{-1cm}I_2=2 (6 a_5-1) p_2^2 q_1 + 4 (1 - 3 a_5) p_3^2 q_1 - 8 a_4^3 (q_2 - \rmi q_3)^4 +
 q_1 (q_2 - \rmi q_3)^2 ((6 a_5-1) q_2 \\
&\hspace*{-1cm}+ 2 \rmi (1 - 3 a_5) q_3)
+
 2 p_1 p_3 (\rmi (q_2 + 6 a_5 q_2) - 2 q_3 + 6 a_5 q_3) +
 32 a_4^2 (p_2 - \rmi p_3) (p_2 q_1 - p_1 q_2 + \rmi p_1 q_3\\
&\hspace*{-1cm}- \rmi p_3 q_1)
+
 2 p_2 (3 \rmi (1 - 4 a_5) p_3 q_1 +
    p_1 (q_2 - 6 a_5 q_2 + 2 \rmi (3 a_5-2) q_3)) +
 2 a_4 (-4 p_1 (p_2 - \rmi p_3) q_1\\
&\hspace*{-1cm} - 6 p_3^2 q_2
+ 4 p_1^2 (q_2 - \rmi q_3) +
    6 \rmi p_2^2 q_3 +
    6 p_2 p_3 (-\rmi q_2 + q_3) + (q_2 - \rmi q_3)^2 (q_1^2 - (q_2 - \rmi q_3) ((6 a_5-1) q_2 \\
&\hspace*{-1cm}+
          2 \rmi (1 - 3 a_5) q_3))),
 \end{split}
\]
\item  $p=-1$, $\lambda=2$
\[
 \begin{split}
I_1=2 (p_2 - \rmi p_3) (p_1 - a_4 p_2 + \rmi a_4 p_3) + q_1 (q_2 - \rmi q_3)^2,
 \end{split}
\]
\item $p=2$, $\lambda=5$
\[
 \begin{split}
I_1&=4 (p_3^2-p_2^2) q_1+4 p_2 (2 \rmi p_3 q_1 + p_1 (q_2 - \rmi q_3)) +
 p_1 p_3 (-4 \rmi q_2 - 4 q_3) \\
&+ (2 q_1 + a_4 (q_2 - \rmi q_3)) (q_2 - \rmi q_3)^3,
 \end{split}
\]
\item  $p=-2$, $\lambda=7$
\[
 \begin{split}
I_1&=(p_2 - \rmi p_3) (2 a_4 (2 (p_2 - \rmi p_3)^2 + 3 (q_2 - \rmi q_3)^3) +
    21 q_1 (q_2 - \rmi q_3)^2)\\
& + 3 p_1 (4 (p_2 - \rmi p_3)^2 - (q_2 - \rmi q_3)^3),
 \end{split}
\]
\item $p=3$, $\lambda=12$
\[
 \begin{split}
&I_1=  8 (p_2 - \rmi p_3)^3 q_1 - 8 p_1 (p_2 - \rmi p_3)^2 (q_2 - \rmi q_3) -
 2 (p_2 - \rmi p_3) (6 q_1\\
& + a_4 (q_2 - \rmi q_3)) (q_2 - \rmi q_3)^3 +
 p_1 (q_2 - \rmi q_3)^4,
 \end{split}
\]
\item  $p=-3$, $\lambda=15$
\[
 \begin{split}
  &I_1=(8 p_1 + 5 a_4 (p_2 - \rmi p_3)) (p_2 - \rmi p_3)^3 +
 30 (p_2 - \rmi p_3)^2 q_1 (q_2 - \rmi q_3)^2\\
& -
 6 (p_2 - \rmi p_3) (p_1 - a_4 p_2 + \rmi a_4 p_3) (q_2 - \rmi q_3)^3 -
 3 q_1 (q_2 - \rmi q_3)^5,
 \end{split}
\]
\item $p=4$, $\lambda=22$
\[
 \begin{split}
&I_1=  112 p_2^4 q_1 - 112 p_1 p_2^3 q_2 + 42 p_1 p_2 q_2^4 -
 28 p_2^2 q_2^3 (12 q_1 + a_4 q_2) + q_2^6 (21 q_1 + 2 a_4 q_2).
 \end{split}
\]
\end{itemize}

If $\lambda$ belongs to the second item of the Morales-Ramis table \eqref{eq:tabMoRa}, i.e.
\[
 \lambda=\dfrac{1}{2}\left(\dfrac{2}{3}+3p(p+1)\right), \qquad p\in\Z,
\]
then potential is super-integrable with one additional first integral and as commuting first integrals we can choose
$\{I_0,I_1\}$ or $\{I_0,I_2\}$.\\
$\blacktriangleright$ $p=0$, $\lambda=1/3$
\[
\begin{split}
&I_1=12 p_1 (p_2 - \rmi p_3) (q_1 + 3 a_4 (q_2 - \rmi q_3)) -
 6 a_4 q_1 (6 (p_2 - \rmi p_3)^2 + (q_2 - \rmi q_3)^3) - 12 p_1^2 (q_2 - \rmi q_3) \\
&+
 q_1^2 (q_2 - \rmi q_3)^2 + 9 a_4^2 (q_2 - \rmi q_3)^4,\\
&I_2=108 a_4 p_1^2 (p_2 - \rmi p_3)-12 p_1^3-3 p_1 (9 a_4^2 (12 (p_2 - \rmi p_3)^2 + (q_2 - \rmi q_3)^3) + q_1^2 (q_2 - \rmi q_3)\\
& -
    6 a_4 q_1 (q_2 - \rmi q_3)^2)
 + (p_2 - \rmi p_3) (q_1^3 +
    54 a_4^3 (6 (p_2 - \rmi p_3)^2 + (q_2 - \rmi q_3)^3) -
    27 a_4^2 q_1 (q_2 - \rmi q_3)^2),
\end{split}
\]
$\blacktriangleright$ $p=1$, $\lambda=10/3$
\[
 \begin{split}
 &I_1= 4410 p_1^3 (p_2 - \rmi p_3) +
 180 p_1 (p_2 - \rmi p_3) (54 a_4^2 ((p_2 - \rmi p_3)^2 + (q_2 - \rmi q_3)^3) +
    49 q_1^2 (q_2 - \rmi q_3)\\
& + 105 a_4 q_1 (q_2 - \rmi q_3)^2) -
 63 p_1^2 (108 a_4 (p_2 - \rmi p_3)^2 - 35 q_1 (q_2 - \rmi q_3)^2 +
    66 a_4 (q_2 - \rmi q_3)^3) \\
&-
 4 (245 q_1^3 (3 (p_2 - \rmi p_3)^2 - 4 (q_2 - \rmi q_3)^3) +
    2430 a_4^3 (p_2 - \rmi p_3)^2 (2 (p_2 - \rmi p_3)^2 + (q_2 - \rmi q_3)^3) \\
&+
    27 a_4^2 q_1 (261 (p_2 - \rmi p_3)^2 + 32 (q_2 - \rmi q_3)^3) (q_2 - \rmi q_3)^2 -
    756 a_4 (9 (-1 + 4 a_5) (p_2 - \rmi p_3)^4 \\
&-
       6 (p_2 - \rmi p_3)^3 (p_2 + \rmi p_3) -
       2 (p_2^2 + p_3^2) (q_2 - \rmi q_3)^3 - (q_2 - \rmi q_3)^5 (q_2 + \rmi q_3) \\
&+
       2 (p_2 - \rmi p_3)^2 (q_2 - \rmi q_3) (-5 q_1^2 +
          3 (q_2 - \rmi q_3) ((2 a_5-1) q_2 - 2 \rmi a_5 q_3)))),\\
&I_2=4 (-42 a_4 q_1 (18 (p_2 - \rmi p_3)^4 + 9 (p_2 - \rmi p_3)^2 (q_2 - \rmi q_3)^3 -
       8 (q_2 - \rmi q_3)^6) +
    49 q_1^2 (-15 (p_2 - \rmi p_3)^2 \\
&+ 4 (q_2 - \rmi q_3)^3) (q_2 - \rmi q_3)^2 +
    9 a_4^2 (21 (p_2 - \rmi p_3)^2 + 16 (q_2 - \rmi q_3)^3) (q_2 - \rmi q_3)^4) \\
&-
 147 p_1^2 (-24 (p_2 - \rmi p_3)^2
+ (q_2 - \rmi q_3)^3) (q_2 - \rmi q_3) +
 84 p_1 (p_2 - \rmi p_3) ((49 q_1 \\
&+ 39 a_4 (q_2 - \rmi q_3)) (q_2 - \rmi q_3)^3
-
    6 (p_2 - \rmi p_3)^2 (7 q_1 - 6 a_4 q_2 + 6 \rmi a_4 q_3)),
 \end{split}
\]
$\blacktriangleright$  $p=-1$, $\lambda=1/3$
\[
 \begin{split}
  &I_1=12 p_1 (p_2 - \rmi p_3) (q_1 + 3 a_4 (q_2 - \rmi q_3)) -
 6 a_4 q_1 (6 (p_2 - \rmi p_3)^2 + (q_2 - \rmi q_3)^3) - 12 p_1^2 (q_2 - \rmi q_3)\\
& +
 q_1^2 (q_2 - \rmi q_3)^2 + 9 a_4^2 (q_2 - \rmi q_3)^4,\\
&I_2=12 p_1^3 - 108 a_4 p_1^2 (p_2 - \rmi p_3) +
 3 p_1 (9 a_4^2 (12 (p_2 - \rmi p_3)^2 + (q_2 - \rmi q_3)^3) + q_1^2 (q_2 - \rmi q_3)\\
&-
    6 a_4 q_1 (q_2 - \rmi q_3)^2)
- (p_2 - \rmi p_3) (q_1^3 +
    54 a_4^3 (6 (p_2 - \rmi p_3)^2 + (q_2 - \rmi q_3)^3) -
    27 a_4^2 q_1 (q_2 - \rmi q_3)^2).
 \end{split}
\]

For the next four items of the Morales-Ramis table~\eqref{eq:tabMoRa}  corresponding potentials are also super-integrable but explicit forms of first integrals are more complicated. Thus we reproduce here explicit expressions for the first element from the family
\[
 \lambda=-\dfrac{1}{24} + \dfrac{1}{6}(1 + 3p)^2, \qquad p\in\Z,
\]
for $p=0$ that gives $\lambda=1/8$ and  the additional  first integrals are
\[
 \begin{split}
  &I_1=31360 p_1^4 - 301056 a_4 p_1^3 (p_2 - \rmi p_3) +
 24 (37888 a_4^4 - 147 (1 - 4 a_5)^2\\
& - 22848 a_4^2 (-1 + 4 a_5)) (p_2 -
    \rmi p_3)^4
+ 672 (-7 + 544 a_4^2 + 28 a_5) (p_2 - \rmi p_3)^3 (p_2 + \rmi p_3) \\
&-
 4 p_1 (p_2 - \rmi p_3) (343 q_1^3 +
    16 a_4 (16 a_4 (16 a_4 (96 (p_2 - \rmi p_3)^2
+ 5 (q_2 - \rmi q_3)^3) -
          63 q_1 (q_2 - \rmi q_3)^2) \\
&+ 147 q_1^2 (q_2 - \rmi q_3))) +
 112 p_1^2 (6 (-7 + 2080 a_4^2 + 28 a_5) (p_2 - \rmi p_3)^2
 -
    28 (p_2^2 + p_3^2) + (35 q_1^2\\
& - 224 a_4 q_1 (q_2 - \rmi q_3) +
       2 (640 a_4^2 (q_2 - \rmi q_3) - 7 (q_2 + \rmi q_3)) (q_2 - \rmi q_3)) (q_2 -
       \rmi q_3)) \\
&-
 28 (p_2^2 + p_3^2) (q_2 - \rmi q_3) (7 q_1^2 + 224 a_4 q_1 (q_2 - \rmi q_3) -
    8 (q_2 - \rmi q_3) ((-7 + 544 a_4^2) q_2 \\
&- \rmi (7 + 544 a_4^2) q_3))
- (q_2 -
    \rmi q_3)^2 (49 q_1^4 +
    64 a_4 q_1 (-3840 a_4^2 (q_2 - \rmi q_3) + 49 (q_2 + \rmi q_3)) (q_2 -
       \rmi q_3)^2 \\
&+
    56 (-1088 a_4^2 (q_2 - \rmi q_3)
+ 7 (q_2 + \rmi q_3)) (q_2 - \rmi q_3)^2 (q_2 +
       \rmi q_3) + 98 q_1^2 (q_2^2 + q_3^2)) \\
&+
 2 (p_2 - \rmi p_3)^2 (-784 (p_2 + \rmi p_3)^2 +
    1568 a_4 q_1 (q_1^2 + 3 (-1 + 4 a_5) (q_2 - \rmi q_3)^2) \\
&+
    340992 a_4^3 q_1 (q_2 - \rmi q_3)^2
+ 40960 a_4^4 (q_2 - \rmi q_3)^3 +
    11424 a_4^2 (q_2 - \rmi q_3) (q_1^2 \\
&+
       8 (q_2 - \rmi q_3) ((2 - 4 a_5) q_2 + 4 \rmi a_5 q_3))
+
    147 (-1 + 4 a_5) (q_2 - \rmi q_3) (q_1^2 + 8 (q_2^2 + q_3^2))),\\
&I_2=263424 p_1^3 (p_2 - \rmi p_3) (7 q_1 + 48 a_4 (q_2 - \rmi q_3)) -
 4704 p_1^2 (224 a_4 q_1 (12 (p_2 - \rmi p_3)^2 + (q_2 - \rmi q_3)^3) \\
&-
    256 a_4^2 (-24 (p_2 - \rmi p_3)^2 + (q_2 - \rmi q_3)^3) (q_2 - \rmi q_3) -
    49 q_1^2 (q_2 - \rmi q_3)^2) \\
&+
 112 p_1 (p_2 -
    \rmi p_3) (5376 a_4^2 q_1 (48 (p_2 - \rmi p_3)^2 + 7 (q_2 - \rmi q_3)^3) -
    343 q_1^3 (q_2 - \rmi q_3) \\
&+
    4096 a_4^3 (48 (p_2 - \rmi p_3)^2
- 11 (q_2 - \rmi q_3)^3) (q_2 - \rmi q_3) -
    7056 a_4 q_1^2 (q_2 - \rmi q_3)^2) \\
&+
 2401 q_1^4 (4 (p_2 - \rmi p_3)^2 + (q_2 - \rmi q_3)^3)
 -
 114688 a_4^3 q_1 (192 p_2^4 - 768 \rmi p_2^3 p_3 + 192 p_3^4\\
& +
    16 \rmi p_2 p_3 (48 p_3^2 - 5 (q_2 - \rmi q_3)^3) -
    8 p_2^2 (144 p_3^2 - 5 (q_2 - \rmi q_3)^3)
 -
    40 p_3^2 (q_2 - \rmi q_3)^3 + (q_2 - \rmi q_3)^6) \\
&-
 1843968 p_1^4 (q_2 - \rmi q_3) +
 75264 a_4^2 q_1^2 (12 (p_2 - \rmi p_3)^2
+ (q_2 - \rmi q_3)^3) (q_2 - \rmi q_3)^2\\
& -
 21952 a_4 q_1^3 (q_2 - \rmi q_3)^4 +
 65536 a_4^4 (84 (p_2 - \rmi p_3)^2 + (q_2 - \rmi q_3)^3) (q_2 - \rmi q_3)^4,
 \end{split}
\]
with $\{I_0,I_1\}=\{I_0,I_2\}=0$.


\newcommand{\noopsort}[1]{}\def\polhk#1{\setbox0=\hbox{#1}{\ooalign{\hidewidth
  \lower1.5ex\hbox{`}\hidewidth\crcr\unhbox0}}} \def\cprime{$'$}
  \def\cydot{\leavevmode\raise.4ex\hbox{.}} \def\cprime{$'$} \def\cprime{$'$}
  \def\polhk#1{\setbox0=\hbox{#1}{\ooalign{\hidewidth
  \lower1.5ex\hbox{`}\hidewidth\crcr\unhbox0}}} \def\cprime{$'$}
  \def\cprime{$'$} \def\cprime{$'$} \def\cprime{$'$} \def\cprime{$'$}
  \def\cprime{$'$}

\end{document}